\documentclass[aps,prl,twocolumn,showpacs,nofootinbib,superscriptaddress]{revtex4-1}
\usepackage{graphicx}
\usepackage{color}
\usepackage{epsfig}
\usepackage{amsmath,amssymb,amsthm}
\usepackage{dcolumn,bm}
\usepackage{color,soul}

\def\[{\left[}
\def\]{\right]}
\def\({\left(}
\def\){\right)}
\newcommand{\iisc}
{\affiliation{Centre for Condensed Matter Theory, Department of Physics, Indian Institute of Science, Bangalore 560012, India}}

\newcommand{\imsc}
{\affiliation{The Institute of Mathematical Sciences, Chennai 600113, India}}

\newcommand{\icts}
{\affiliation{International Centre for Theoretical Sciences, TIFR, Bangalore 560089, India}}

\begin{document}
\title
{Thermal Conductivity of Glass-Forming Liquids}


\author{Pranab Jyoti Bhuyan}
\email[Email: ]{pranab@physics.iisc.ernet.in}
\iisc

\author{Rituparno Mandal}
\email[Email: ]{rituparno@physics.iisc.ernet.in}
\iisc

\author{Pinaki Chaudhuri}
\email[Email: ]{pinakic@imsc.res.in }
\imsc

\author{Abhishek Dhar}
\email[Email: ]{abhishek.dhar@icts.res.in }
\icts

\author{Chandan Dasgupta}
\email[Email: ]{cdgupta@physics.iisc.ernet.in}
\iisc

\begin{abstract}

Thermal conductivity of a model glass-forming system in the liquid and glass states is studied using extensive numerical simulations. We show that near the glass transition temperture, where the structural relaxation time becomes very long, the measured thermal conductivity decreases with increasing age. Secondly the thermal conductivity of the disordered solid obtained at low temperatures depends on the cooling rate with which it was prepared, with lower cooling rates leading to lower thermal conductivity. Our analysis links this decrease of the thermal conductivity with increased exploration of lower-energy inherent structures of the underlying potential energy landscape. Further we show that the lowering of conductivity for lower-energy inherent structures is related to the high frequency harmonic modes associated with the inherent structure
being less extended.

\end{abstract}

\pacs{61.20.Ja,63.50.Lm, 44.10.+i}
\maketitle

{\em Introduction} -- Thermal conductivity is an important material property with  wide ranging applications and understanding the mechanism of heat transport in different materials is crucial from both theoretical and applied perspectives. It is well known that crystalline materials transport heat more efficiently than glassy and disordered materials, for which the thermal conductivity ($\kappa$) can be orders of magnitude smaller \cite{Cahill:88,Pohl:71}.

A well-studied feature of glassy systems is the temperature $(T)$-dependence of their thermal conductivity, which is very distinct from crystals. One typically finds $\kappa \sim T^2$ at very low temperatures, followed by a pronounced plateau and then an eventual gradual increase with temperature.   The low temperature
features are expected to be of quantum-menchanical origin and a range of mechanisms have been proposed to explain them \cite{Anderson:72,Buchenau:92,Clare:87}. One scenario relates the plateaue to the presence of excess modes, the so-called {\it boson peak}, in the low-frequency regime of the vibrational spectrum of glasses \cite{Lubchenko:03,schirmacher:2006,Nagel:10}, which has thereafter been connected with the presence of elastic heterogeneities in such materials \cite {Maruzzo:2013,Barrat:16}.

More recently,  a number of studies have looked at higher temperatures where classical physics dominates. This includes regimes much below the glass transition temperature where the system gets stuck in the basin of a low-energy {\it inherent structure} (IS) (local minimum of the potential energy), as well as intermediate temperatures where the system evolves slowly while undergoing transitions between different basins. In the former case the system can be effectively described by a disordered harmonic model obtained by expanding the many-body potential about the  minimum. Heat transport is then related to the diffusivity associated with the normal modes of the harmonic solid \cite{Allen:89,Allen:93}. In the context of the jamming transition, the change in the form of the diffusivity in a soft-sphere packing was studied in \cite{Nagel:09,Nagel:10,wyart:10}, and a crossover was observed near the boson-peak frequency as the transition was approached. In \cite{Barrat:13}, it was shown that the introduction of polydispersity leads to an amorphisation transition, whereby the thermal conductivity of the system decreases considerably.

So far the question as to whether thermal transport in glasses depends upon the age of the material, or on the cooling rate used in its preparation has not been explored. Here we address this question. It is well-known that the material properties of glasses depend on their age and history of preparation, e.g. the cooling rate by which they were quenched from a fluid phase \cite{Kob:97, Binder:96, glassbook}. This dependence on the history can be seen, for example, in the transient response of glasses to applied shear \cite{Shi:06, Moorcroft:11, Chaudhuri:16}. Surprisingly, in the case of thermal transport, there is not much systematic study. Using extensive numerical simulations of a model structural glass former, we show that the thermal conductivity does depend on the age of the glass and on the rate of cooling during preparation.  Further, we demonstrate that the decrease of thermal conductivity in glasses with growing age or slower cooling rate is linked to the exploration of ISs with lower energy. By calculating the participation ratio of the normal modes associated with different ISs, we see that the decrease in $\kappa$ is related to an increase in the degree of localization of high frequency vibrational modes. 

Another  puzzling  issue that we address relates to the question of whether the thermal conductivity of the disordered harmonic system that describes the dynamics of small displacements from an IS is infinite. Rayleigh scattering of low frequency ($\omega$) phonons is expected to lead to mean free paths $\ell \sim 1/\omega^4$ \cite{ziman}, which can cause a divergence of $\kappa$ and this is seen in some recent studies \cite{Nagel:09,kundu:10,chaudhuri:10}. In our study we compute the thermal conductivity $\kappa$ of a glass former using three numerical methods (see below), all of which indicate a finite conductivity as $T \to 0$, where the harmonic approximation should be valid. However, we can not rule out a divergence because the system sizes used in our simulations are modest (number of particles $\leq 20,000$). A recent study~\cite{lerner2016} has shown that the low-frequency part of the vibrational spectrum for such values of $N$ is dominated by the \textit{boson peak} modes and the effects of long-wavelength phonon modes are observed only in systems with much larger values of $N$. Since the divergence of $\kappa$ is expected to arise from these phonon modes, simulations of much larger systems would be required to provide a clear answer to the question of whether $\kappa$ diverges as $T\to 0$. Our work shows that the contribution of the boson-peak modes to the thermal conductivity remains finite as $T \to 0$.

We consider the well-known model glass former - the Kob-Andersen binary Lennard-Jones mixture \cite{Kob:94} (details in section I of \textit{Supplementary Information} (SI)). We consider $N$ particles in a box of dimensions $L_x \times L_y \times L_z$. Our study is done for a number density of $\rho=N/(L_x L_y L_z)=1.2$, for which the Vogel-Fulcher-Tamman ($VFT$) transition temperature of this glass former is $T_0 \approx 0.3$ in reduced Lennard-Jones units. We compute  the thermal conductivity at different temperatures above and below the glass transition. The supercooled liquid and glassy states we have studied were obtained via a well-defined cooling protocol~\cite{Sastry:98}. The system is well-equilibrated at a high temperature $(T=2.50)$ in its liquid state and then  cooled to a low temperature through a number of intermediate temperature
steps. The number of intermediate steps and the amount of time spent in each step determine the cooling rate~\cite{Sastry:98}.

{\em Thermal conductivity} -- Denoting the position and the velocity of the $l$-th particle by ${\vec r}_l=\left\{r^\alpha_l\right\}$ and ${\vec v}_l= \left\{v^\alpha_l\right\}$ respectively (where $\alpha=x,y,z$), the energy current density at point ${\vec r}$ is given by \cite{Hansen:06}:
\begin{equation}
  \mathcal{J}^\alpha({\vec r},t)=  \sum_l \delta({\vec r}-{\vec r}_l)[\epsilon_l v_l^\alpha + \frac{1}{2} \sum_{n \neq l}  (r_l^\alpha-r_n^\alpha)j_{l,n}]
 \label{curr_defn}
\end{equation} 
where $\epsilon_l= {\mathbf{v}_l^2}/2+ (1/2) \sum_{n \neq l} V(r_{ln})$ is the energy of the $l$-th particle, and  $ j_{l,n}=\frac{1}{2} \sum_\nu  (v_l^\nu +v_n^\nu )f_{ln}^\nu $, with  $f_{ln}^\alpha=-\partial{V(r_{ln})}/\partial{r_l^\alpha}$. We evaluated the thermal conductivity  by several methods, which we now describe.
 
\begin{figure}
\includegraphics[height = 0.67\linewidth]{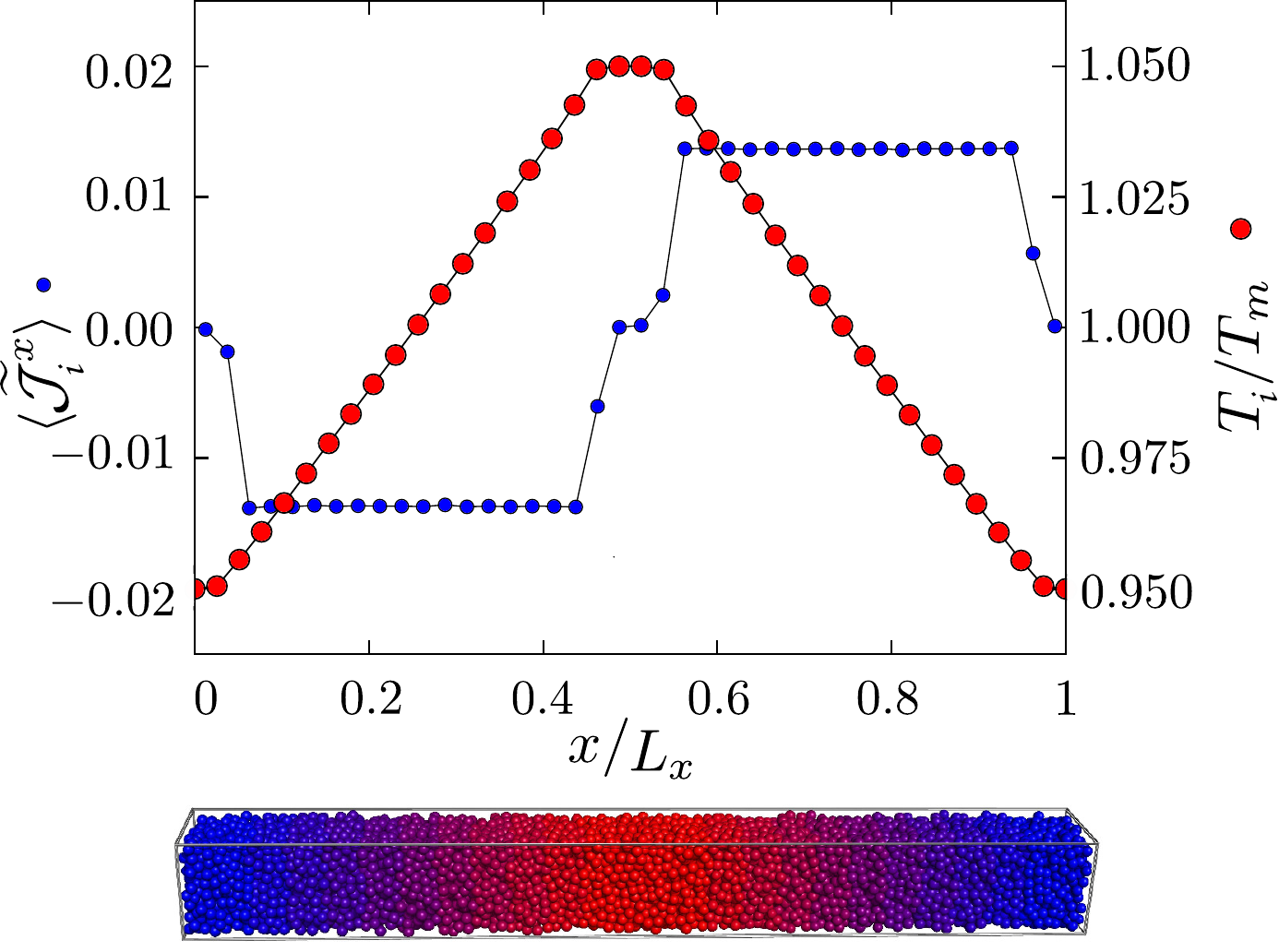}
\caption{(color online). (Top)  Representative plot of local heat current density ($\langle \widetilde{\mathcal{J}}^x_i\rangle $) and reduced temperature ($T_i/T_m$) profile along the system, where $T_m$ is the mean temperature of the sample in a non-equilibrium simulation and (bottom) schematic diagram of the non-equilibrium simulation system, where the color red represents the hotter region and blue represents the colder region. This simulation used $N=10000$ particles in a box with $L_y=L_z=9.41$ and $L_x=10\times L_z$.
}
\label{figure_1}
\end{figure}

{\em Nonequilibrium method (NEMD)} -- We consider a system with mean temperature $T_m$, impose a small  temperature difference $\Delta T$ in the $x$-direction across length $L_x^\prime < L_x$, and measure  the average steady-state heat current density $\langle \mathcal{J}^x \rangle = \frac {1} {V} \int_V d{\vec r} \langle \mathcal{J}^x({\vec r},t)\rangle$ where $V$ is the volume of the region in the sample where the measurement is made. The thermal conductivity is then given by
\begin{equation}
\kappa=\frac{\langle \mathcal{J}^x \rangle \times L^\prime_x}{\Delta T}.
~\label{kappa}
\end{equation}

We consider a rectangular geometry, as illustrated in the schematic diagram of Fig.~\ref{figure_1}. Periodic boundary conditions were maintained in all directions. We first prepare the initial equilibrium states (at mean temperature $T_m$) via the cooling protocol discussed above. A temperature gradient was then applied by keeping a heat source at temperature  $1.05 T_m$ in a region of width $L_x/10$ in the middle of the system, and  two heat sinks  of width $L_x/20$ at the two ends  at $0.95 T_m$. The thermostating of the hot and cold reservoirs was done by drawing the velocities of the particles in each reservoir from a Maxwell-Boltzmann distribution at the appropriate temperatures, at intervals of time 0.1. After an initial transient period ($\delta{t} \sim 10^4$) following the imposition of the thermal gradient, the system reaches a non-equilibrium steady state, wherein the spatial profiles of local temperature and heat current density typically look like those shown in Fig.~\ref{figure_1}. The spatial variation of the local temperature $T_i$ and the heat current density $\langle \widetilde{\mathcal{J}}^x_i\rangle$ was obtained by binning $L_x$ into $40$ segments, labelled $i=1,\ldots40$, and doing the averages separately for each segment with $\delta{t} \sim 5\times{10^4}$, after the transient time (details in section II(a) of SI). Using the steady state current value from simulations Eq.~(\ref{kappa}) with $L^\prime_x=0.4\times L_x$ gives  $\kappa$ at  temperature $T_m$. We have done averages over $32$ independent MD trajectories.

\begin{figure}
\includegraphics[height = 0.62\linewidth]{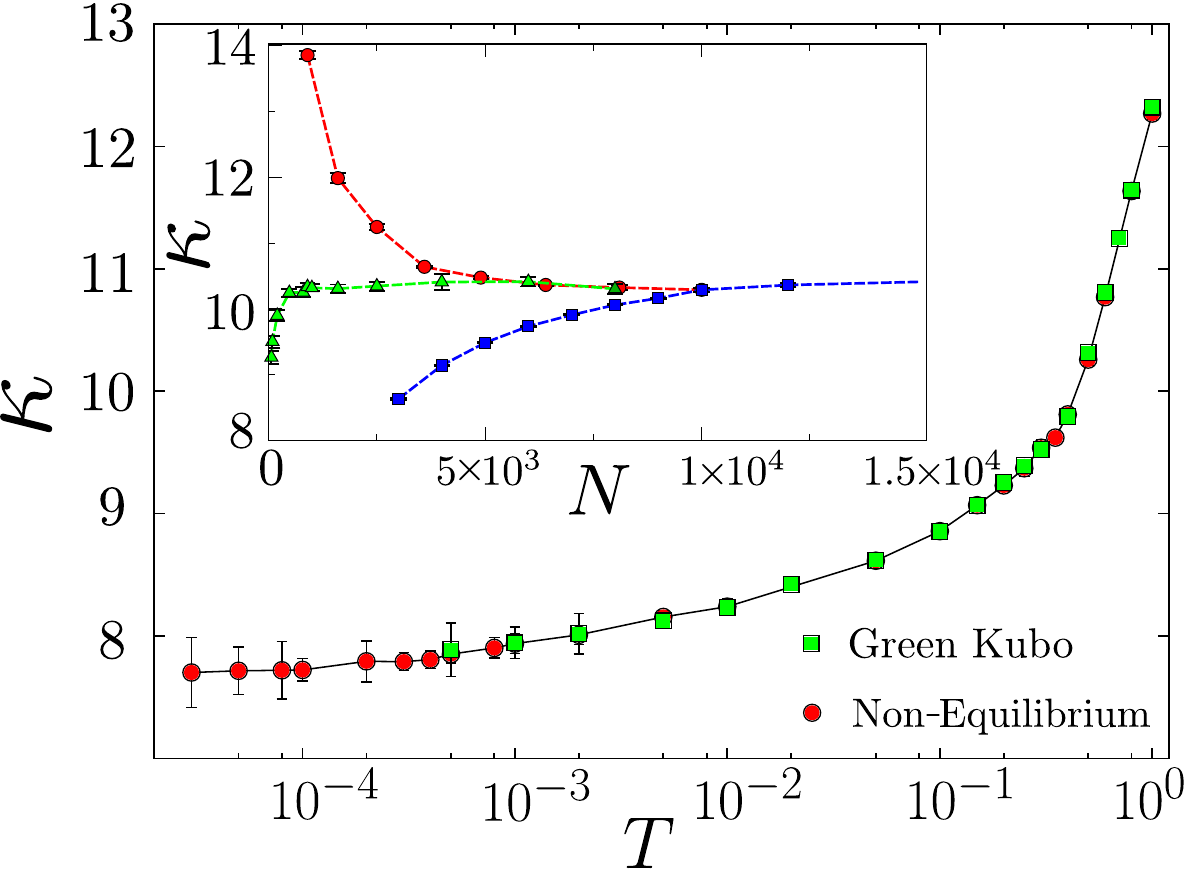}
\caption{(color online). Comparison of thermal conductivity ($\kappa$) values obtained from NEMD simulation and EMD Green-Kubo calculation for a wide range of temperatures for the samples prepared with a cooling rate of $3.3\times{10^{-6}}$. (Inset) $N$ dependence of $\kappa$: NEMD with varying $L_x$ (blue), varying $L_y(=L_z)$ (red), and EMD with varying $L_x(=L_y=L_z)$ (green). (For NEMD $T \equiv T_m$).} 
\label{figure_2}
\end{figure}

{\em Green-Kubo method(EMD)} -- Alternatively, $\kappa$ can be calculated using the Green-Kubo relation \cite{Hansen:06}:
\begin{equation}
\kappa = \frac{1}{3k_BT^2}  \lim_{\tau \to\infty} \lim_{V \to\infty} \frac{\rho}{N} \int_{0}^{\tau} dt \mspace{1mu} \langle {\vec{\mathcal{J}}}_{tot}(0) \cdotp {\vec{\mathcal{J}}}_{tot}(t)\rangle 
\label{gkformula}
\end{equation}
where ${\vec{\mathcal{J}}}_{tot}(t)$ is the integral of the heat current density ${\vec{\mathcal{J}}}({\vec r},t)$ over the whole system and the time correlation function is evaluated at equilibrium. 
\begin{figure}
\includegraphics[height = 0.6\linewidth]{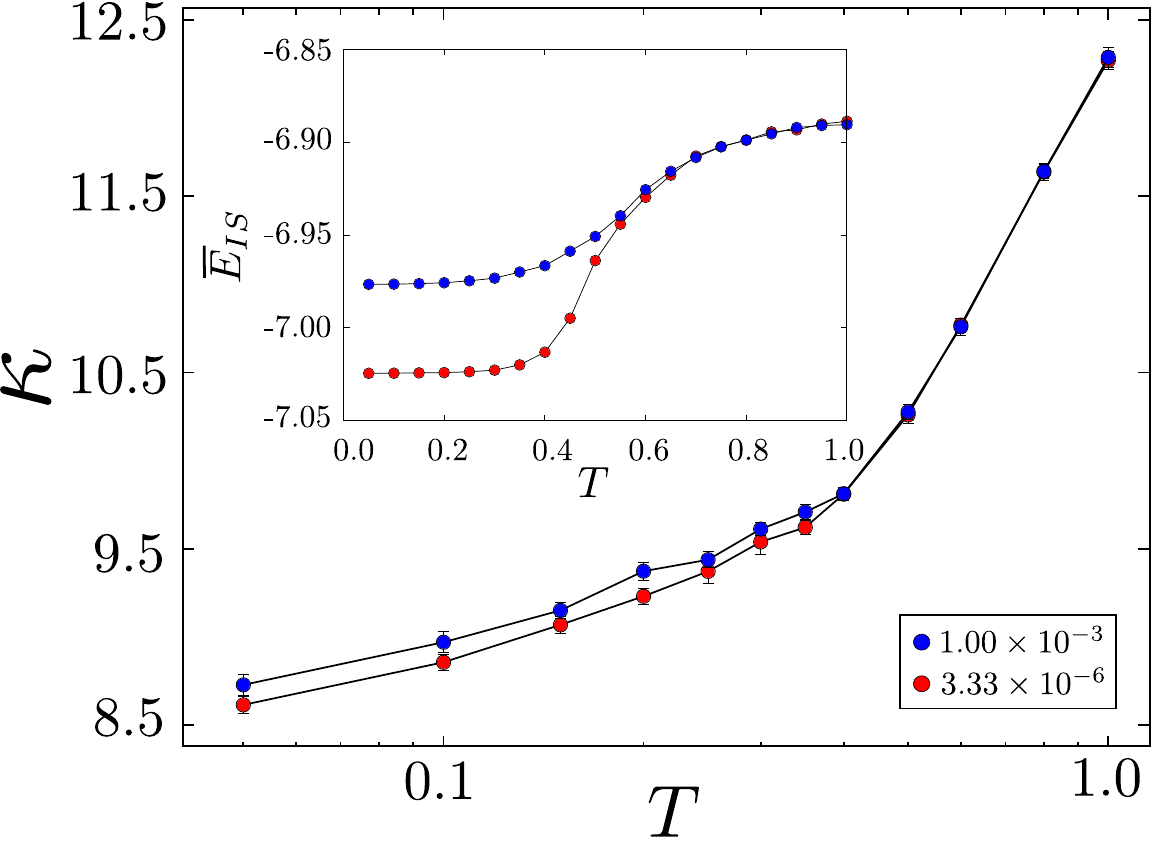}
\caption{(color online). Thermal conductivity ($\kappa$) from NEMD as a function of temperature $T(=T_m)$ for samples prepared using two different cooling rates shows branching as the system approaches the glass transition. The inset shows the mean energy of ISs ($\overline{E}_{IS}$) sampled at $T$ for these cooling rates. The system has N=10000 particles with $L_y=L_z=9.41$ and $L_x=10\times L_z$.}
\label{figure_3}
\end{figure}
For the Green-Kubo calculation we  considered a cubic simulation box, with the same density of particles as in the nonequilibrium case. An initial equilibrium state is prepared following the cooling protocol described earlier, and subsequently the NVE dynamics is switched on. The fluctuations of the heat current are then measured over a time-window of ${\delta}t~\sim10^4$ as the system evolves and the thermal conductivity ($\kappa$) is calculated using the Green-Kubo formula (Eq.~\ref{gkformula}). The estimates of $\kappa$ were obtained via averages over $96$ independent trajectories. 

In Fig.~\ref{figure_2} we compare the values of $\kappa$ measured via the two schemes outlined above, for a given cooling rate.  As expected for a disordered system, the magnitude of the thermal conductivity  continuously decreases over the entire temperature range, before nearly saturating at very low temperatures. We observe very good agreement between the two different measurements of $\kappa$ over a very broad range of temperatures. Surprisingly, this is true even in the glassy regime, where the two methods provide the same result for states which have the same preparation history, i.e. produced via the same cooling rate. In the rest of the work we have used these two methods independently, in accordance with the question to be addressed. By performing simulations at various system sizes  (keeping $\rho$ constant),  we have verified  that the results for $\kappa$ become essentially independent of system size for $N\sim 10^3$ (EMD) and $N\sim 10^4$ (NEMD). In the inset of Fig.~\ref{figure_2} we show the system size dependence of $\kappa$ (details in section II of SI).

{\em Effect of cooling rate} -- It is known that the cooling rate influences the glassy state obtained at low temperatures \cite{glassbook}. For example, the mean energy of the underlying ISs, $\overline{E}_{IS}$, corresponding to any temperature, depends on the choice of the cooling rate \cite{Sastry:98}, with slower cooling leading to exploration of lower energy ISs. Using data from our simulations, this is illustrated in the inset of Fig.~\ref{figure_3}, where we see that below a certain temperature, the variation of  $\overline{E}_{IS}$ with $T$ branches out for two contrasting cooling rates. In the same spirit, we check how $\kappa$ (from non-equilibrium measurements) varies with temperatures along the same two cooling branches. This is shown in  Fig.~\ref{figure_3}, and we observe that the slower cooling corresponds to lower values of $\kappa$. However, compared to data shown for $\overline{E}_{IS}$, in this case, the branching occurs
at a lower temperature, typically where the dynamics is known to fall out of equilibrium \cite{glassbook}.

\begin{figure}
\includegraphics[height = 0.62\linewidth]{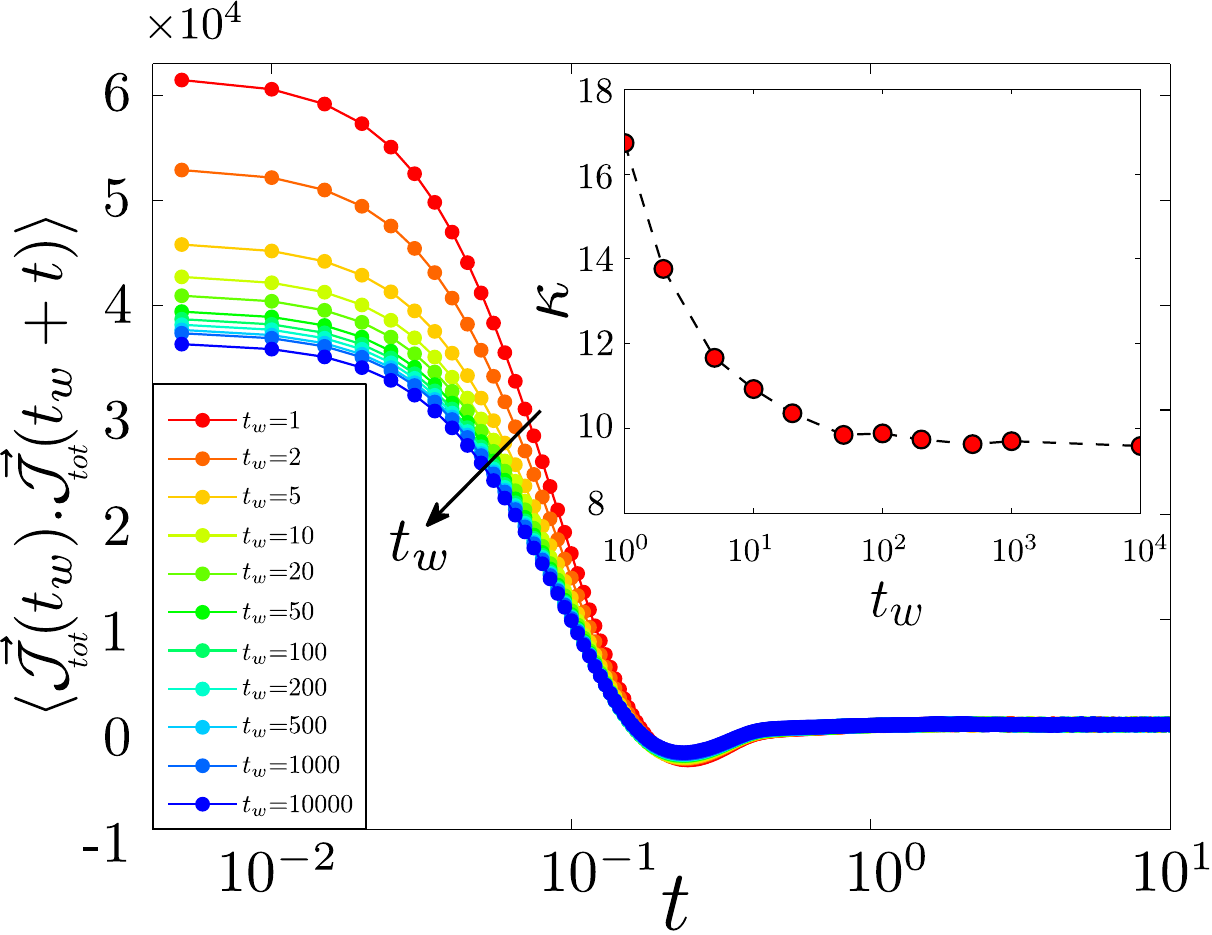}
\caption{(color online). Heat current time auto-correlation function $\langle {\vec{\mathcal{J}}}_{tot}(t_w) \cdotp  {\vec{\mathcal{J}}}_{tot}(t_w+t)\rangle$ with changing age, $t_w$, for a system of $N=1000$ particles with $L_x=L_y=L_z=9.41$ quenched from $T=2.50$ to $T=0.30$. The inset shows the values of thermal conductivity($\kappa$) for different $t_w$.} 
\label{figure_4}
\end{figure}

{\em Effect of aging} -- The dependence of thermal conductivity on the history of preparation of the glassy state, as demonstrated above, implies that $\kappa$ would also vary with the age of the glass when the system evolves after a thermal  quench from high temperature to low temperatures. To investigate this, we quench the system from $T=2.50$ to $T=0.30$ (which is close to the $VFT$ transition temperature) and  let the system evolve in a constant temperature environment  for a time  $t_w$.  Once the system has reached the age $t_w$, the $NVE$ dynamics is switched on. Under such conditions, we measure the total heat current and calculate the corresponding auto-correlation function $\langle{\vec{\mathcal{J}}}_{tot}(t_w) \cdotp \vec{{\mathcal{J}}}_{tot}(t_w+t)\rangle$. In Fig.~\ref{figure_4}, we show the current auto-correlation function for increasing age $t_w$. Note that the correlators relax quickly in time, even though the structural relaxations are extremely slow at this temperature \cite{Scheidler:01}. The variation of conductivity $\kappa$ with changing age $t_w$, calculated from these correlation functions using the Green-Kubo formula, is shown in the inset of Fig.~\ref{figure_4}: with increasing age, the thermal conductivity decreases. Now, it is known that glassy systems explore deeper minima in the potential energy landscape with increasing age \cite{Sciortino:99}. Therefore, the observed dependence of $\kappa$ on $t_w$ and our earlier observation regarding the dependence of $\kappa$ on cooling rate suggest that the thermal conductivity of the glass is closely related to the energy of the relevant ISs.

\begin{figure}
\includegraphics[height = 0.62\linewidth]{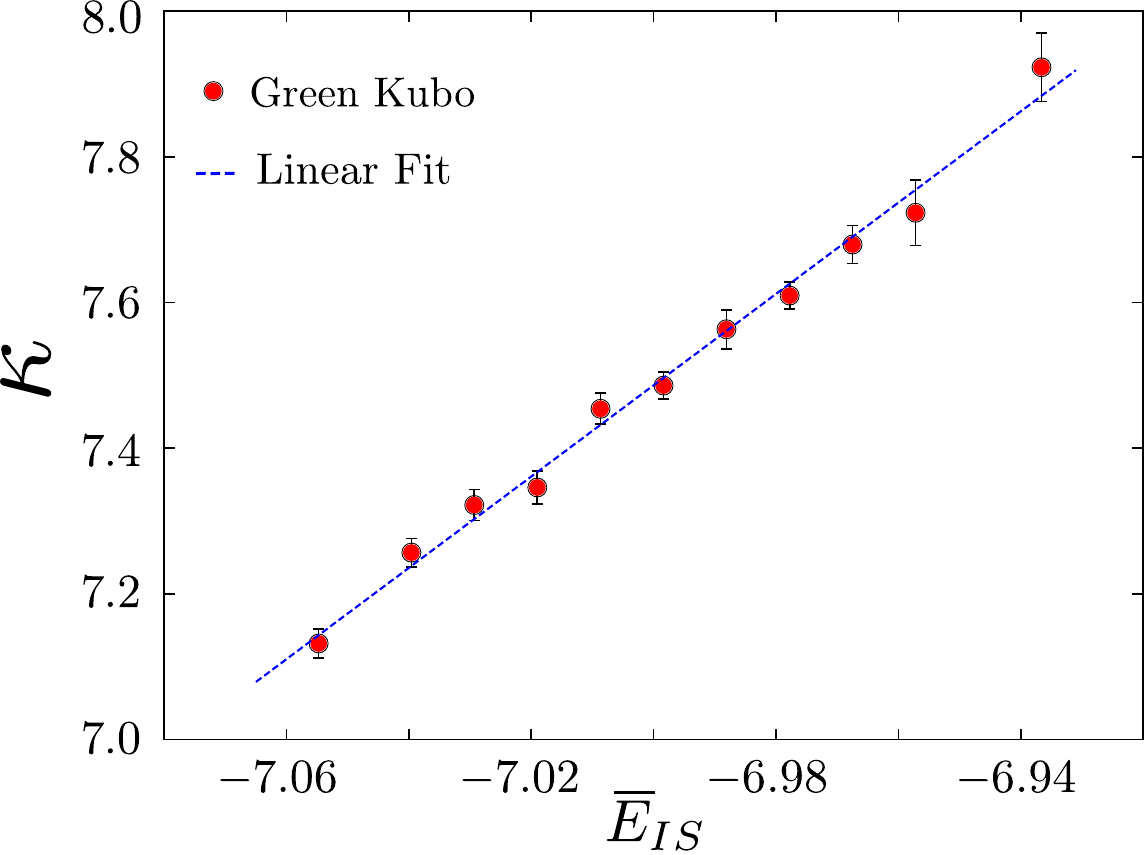}
\caption{(color online). Thermal conductivity ($\kappa$) as a function of mean value of the IS energy ($\overline{E}_{IS}$) from low temperature Green-Kubo simulation starting with IS configurations. The dotted line is a linear fit to the simulation data.}
\label{figure_5}
\end{figure}

{\em Dependence on $E_{IS}$} --  We now explore this possible link between properties of ISs and the values of $\kappa$. We consider a number of ISs with mean energy $\overline{E}_{IS}$ and standard deviation $\sim10^{-6}$ and prepare  low temperature ($T=0.002$) initial states consistent with small harmonic fluctuations around the potential minimum (details in section IV of SI). In the harmonic theory, the total energy of an equilibrium state with temperature $T$ is $\overline{E}_{IS}+[(6N-3) k_B T/2]$. An initial state with this energy  was prepared  by choosing positions corresponding to the IS and giving each particle a random velocity taken from a Maxwell distribution at temperature $2T$. MD simulations within the \textit{NVE} ensemble are carried out for $N=1000$. For the states that remain confined to the basin of the initial IS during the course of the simulation, the thermal conductivity is calculated using the Green-Kubo method.  In Fig.~\ref{figure_5} we show the dependence of $\kappa$ on the  energy of the IS -- we find $\kappa$ increases linearly with $\overline{E}_{IS}$. This explains why the thermal conductivity decreases with slower cooling, or with increasing age of the system after the thermal quench -- the system starts exploring ISs with lower energy. 

{\em Properties of harmonic excitations} -- At very low temperatures, the system mostly remains close to an IS of the potential energy landscape and one can approximate its potential energy to that of a harmonic solid. It is of interest to relate the thermal conductivity to the properties of the harmonic excitations, especially the effect of Anderson localization \cite{nagel84,laird91,kundu:10}. An expansion of the potential energy to quadratic order in the displacements from the IS gives the Hessian matrix  which completely describes the properties of the 
harmonic solid. We ask how the properties of the eigenvalues and eigenfunctions of the Hessian matrix affect thermal transport and the relation between $\kappa$ and $\overline{E}_{IS}$. Following the formulation in \cite{Allen:93}, the conductivity is given by $\kappa= (k_B/V) \int_0^\infty d \omega g(\omega) d(\omega)$, where the density of states $g(\omega)$ and the {\emph{thermal diffusivity}} $d(\omega)$ are known completely in terms  of the eigenvalues and eigenfunctions of the  Hessian matrix (details in section III of SI). We expect the diffusivity of modes to depend on their degree of delocalization which can be quantified  by the participation ratio ($PR(\omega)$).

\begin{figure}
\includegraphics[width=7.1cm]{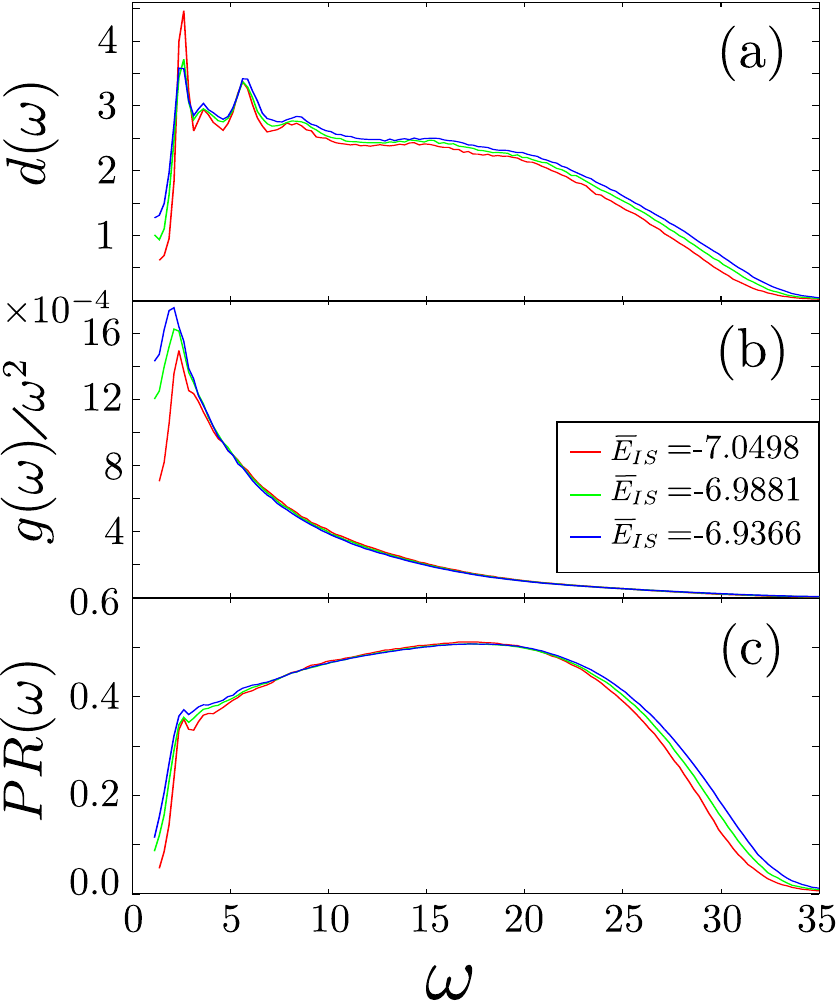} 
\caption{(color online). (a) Heat diffusivity $d(\omega)$ for three different groups of ISs with different $\overline{E}_{IS}$ and (b) Reduced density of states $g(\omega)/\omega^2$ showing the presence of more extra modes near the boson peak with increase in $\overline{E}_{IS}$. (c) Participation ratio for these three groups of ISs showing the presence of more localized low and high frequency modes with decrease in $\overline{E}_{IS}$.}
\label{figure_6}
\end{figure}

In Fig.~\ref{figure_6}(a), we consider ISs with three different $\overline{E}_{IS}$ and compare the results for $d(\omega), g(\omega)$ and $PR(\omega)$ for the three cases. We see that, $d(\omega)$ of the ISs with lower $\overline{E}_{IS}$ are smaller over most of the allowed frequency range except for a small region  near  the boson peak, as compared to the ISs with higher $\overline{E}_{IS}$. We recall that for amorphous solids, excess density of states compared to the Debye law, $g(\omega) \sim \omega^2$, appears in the low frequency region, and this is known as the boson peak \cite{glassbook}. Here we find that the boson peak is more pronounced for ISs with higher $\overline{E}_{IS}$  than the ISs with lower $\overline{E}_{IS}$, as shown in Fig.~\ref{figure_6}(b). Finally from the PR (shown in Fig.~\ref{figure_6}(c)) we see that both the low  and high frequency vibrational modes are more extended for the ISs with higher $\overline{E}_{IS}$. This
indicates that the higher thermal conductivity is related to the increased delocalization of these vibrational modes, with the high frequency modes playing a more important role (details in section III(e) of SI).

{\em Conclusions} --- Using numerical simulations, we have studied thermal transport in a model glass-forming liquid. The values of the thermal conductivity, obtained from non-equilibrium and Green-Kubo calculations are found to agree well over a wide range of temperatures. One of our main findings is that the phenomenon of \textit{aging} has a significant effect on thermal transport. Near the glass transition temperature, the conductivity  drops by about $50\%$ with increasing age. Similarly the conductivity of the low temperature glass can vary by about $10\%$ depending on the coooling protocol used to form it. The lowering of conductivity with growing age or slower cooling can be rationalized in terms of the system exploring lower  energy minima  and our numerical results demonstrating that the thermal conductivity increases linearly with the energy of the IS. Further analysis of the harmonic solid associated with the ISs helped us to understand the underlying reasons in terms of the diffusivity and the extent of localization of the normal modes. For the disordered solid, we do not find a diverging diffusivity at low frequencies, but we can not rule out the possibility of a divergence in much larger systems for which long-wavelength phonon modes are expected to dominate the low-frequency part of the vibrational spectrum.  

\vspace{0.7cm}
We  thank J.-L. Barrat and J. Horbach for discussions;  D. Banerjee and H. Khare for their help with figures. PJB \& RM acknowledges financial support from CSIR, India. We thank SERC IISc, IMSc and ICTS for computing support. AD thanks  support   from the Indo-Israel joint research project No. 6-8/2014(IC) and from the  French Ministry of Education through the grant ANR (EDNHS). CD acknowledges support from DST, India.



\begin{thebibliography}{99}
\bibitem{Cahill:88} D. G. Cahill and R. O. Pohl, Annu. Rev. Phys. Chem. \textbf{39}, 93 (1988). 
\bibitem{Pohl:71} R. C. Zeller and R. O. Pohl, Phys. Rev. B. \textbf{4}, 6 (1971).
\bibitem{Anderson:72} P. W. Anderson, B. I. Halperin, and C. M. Varma, Philos. Mag. \textbf{25}, 1 (1972).
\bibitem{Clare:87}  C. C. Wu, and J. J. Freeman,  Phys. Rev. B {\bf 36}, 7620 (1987). 
\bibitem{Buchenau:92} U. Buchenau, Yu. M. Galperin, V. L. Gurevich, D. A. Parshin, M. A. Ramos, and H, R. Schober, Phys. Rev. B \textbf{46}, 2798 (1992).
\bibitem{Lubchenko:03} V. Lubchenko and P. G. Wolynes, Proc. Natl. Acad. Sci. U.S.A. \textbf{100}, 1515 (2003).
\bibitem{schirmacher:2006} W. Schirmacher, Europhys. Lett. {\bf 73},  892 (2006).
\bibitem{Nagel:10} V. Vitelli, N. Xu, M. Wyart, A. J. Liu and S. R. Nagel, Phys. Rev. E. \textbf{81}, 021301 (2010).
\bibitem{Maruzzo:2013} A. Marruzzo, W. Schirmacher, A. Fratalocchi A and G. Ruocco, Sci. Rep., 3  1407 (2013).
\bibitem{Barrat:16} H. Mizuno, S. Mossa, and J.L. Barrat, Phys. Rev.
B {\bf 94}, 144303 (2016).
\bibitem{Allen:89} P. B. Allen and J. L. Feldman, Phys. Rev. Lett. \textbf{62}, 645 (1989).
\bibitem{Allen:93} P. B. Allen and J. L. Feldman, Phys. Rev. B. \textbf{48}, 12581 (1993).
\bibitem{Nagel:09} N. Xu, V. Vitelli, M. Wyart, A. J. Liu and S. R. Nagel, Phys. Rev. Lett. \textbf{102}, 038001 (2009).
\bibitem{wyart:10} M. Wyart, Europhys. Lett. {\bf 89}, 64001 (2010).
\bibitem{Barrat:13} H. Mizuno, S. Mossa, and J.L. Barrat, Europhys. Lett. \textbf{104}, 56001 (2013)
\bibitem{Kob:97} W Kob, JL Barrat, Phys. Rev. Lett. 78, 4581 (1997).
\bibitem{Binder:96} K. Vollmayr, W. Kob and K. Binder, J. Chem. Phys. \textbf{105}, 4714 (1996).
\bibitem{glassbook} K. Binder and W. Kob, {\it Glassy Materials and Disordered Solids: An Introduction to Their Statistical Mechanics} (World Scientific, Singapore, 2011).
\bibitem{Shi:06}  Y. Shi and M. L. Falk,  {Phys. Rev. B} {\bf 73}, 214201 (2006).
\bibitem{Moorcroft:11} R. L. Moorcroft, M. E. Cates, and S. M. Fielding,  {Phys. Rev. Lett.} {\bf 106}, 055502 (2011).
\bibitem{Chaudhuri:16} GP Shrivastav, P Chaudhuri, J Horbach, Journal of Rheology 60 (5), 835 (2016).
\bibitem{ziman} J. M. Ziman, Principles of the Theory of Solids (Cambridge University Press), Cambridge, 1972.
\bibitem{kundu:10} A. Kundu, A. Chaudhuri, D. Roy, A. Dhar, J. L. Lebowitz, H. Spohn, Europhys. Lett. {\bf 90}, 40001 (2010).
\bibitem{chaudhuri:10} A. Chaudhuri, A. Kundu, D. Roy, A. Dhar, J.L. Lebowitz, H. Spohn,   Phys. Rev. B {\bf 81}, 064301 (2010).
\bibitem{lerner2016} E. Lerner, G. D\"uring, and E. Bouchbinder, Phys. Rev. Lett. {\bf 117}, 035501 (2016).
\bibitem{Kob:94} W. Kob and H. C. Andersen, Phys. Rev. Lett. \textbf{73}, 1376 (1994).
\bibitem{Sastry:98} S. Sastry, P. G. Debenedetti, F. H. Stillinger, Nature \textbf{393}, 554 (1998)
\bibitem{Hansen:06} J. P. Hansen and I. R. McDonald, Theory of Simple Liquids, 3rd ed. (Academic Press, London, 2006).
\bibitem{Scheidler:01} P Scheidler, W Kob, A Latz, J Horbach, K Binder, Phys. Rev. B 63, 104204 (2001).
\bibitem{Sciortino:99} F Sciortino, W Kob, P Tartaglia, Phys. Rev. Lett. 83, 3214 (1999).
\bibitem{nagel84} S.R. Nagel, G.S. Grest, A. Rahman, Phys. Rev. Lett. {\bf 53}, 368 (1984).
\bibitem{laird91} B. B. Laird and H. R. Schober, Phys. Rev. Lett.  {\bf 66}, 636 (1991).

\end{thebibliography}
\end{document}


\title
{Thermal Conductivity of Glass-Forming Liquids - Supplementary Information}


\author{Pranab Jyoti Bhuyan}
\email[Email: ]{pranab@physics.iisc.ernet.in}
\iisc

\author{Rituparno Mandal}
\email[Email: ]{rituparno@physics.iisc.ernet.in}
\iisc

\author{Pinaki Chaudhuri}
\email[Email: ]{pinakic@imsc.res.in }
\imsc

\author{Abhishek Dhar}
\email[Email: ]{abhishek.dhar@icts.res.in }
\icts

\author{Chandan Dasgupta}
\email[Email: ]{cdgupta@physics.iisc.ernet.in}
\iisc

\maketitle
\section{I. Details of model}
For our study, we consider the well-studied Kob-Andersen binary Lennard Jones mixture, which is a model glass former ~\cite{Kob:94}, consisting of $80:20$ proportion of A-type and B-type particles interacting via the pair potential of the form 
\begin{equation} V_{ij}(r)=4 \epsilon_{ij}
\left[\left(\frac{\sigma_{ij}}{r}\right)^{12}-\left(\frac{\sigma_{ij}}{r}\right)^{6}\right]~,
\label{eqn.Potential}
\end{equation}
where $r=r_{ij}$ is the distance between the $i$-th and the $j$-th particle and the indices $i$, $j$ can be A or B. The values of $\sigma_{ij}$ and $\epsilon_{ij}$ are chosen to be: $\sigma_{AB}=0.8 \sigma_{AA}$, $\sigma_{BB}=0.88 \sigma_{AA}$, $\epsilon_{AB}=1.5 \epsilon_{AA}$, $\epsilon_{BB}=0.5 \epsilon_{AA}$. The potential is cut off at $r^c_{ij}=2.5 \sigma_{ij}$ and shifted accordingly using a smoothing function. The units of length and energy are set by  $\sigma_{AA}=1$ and $\epsilon_{AA}=1$. All particles are of unit mass. The numerical integration of the equations of motion are done using the velocity-Verlet algorithm.


\section{II. Finite Size Effects}
\begin{figure}[ht]
\includegraphics[scale = 0.30]{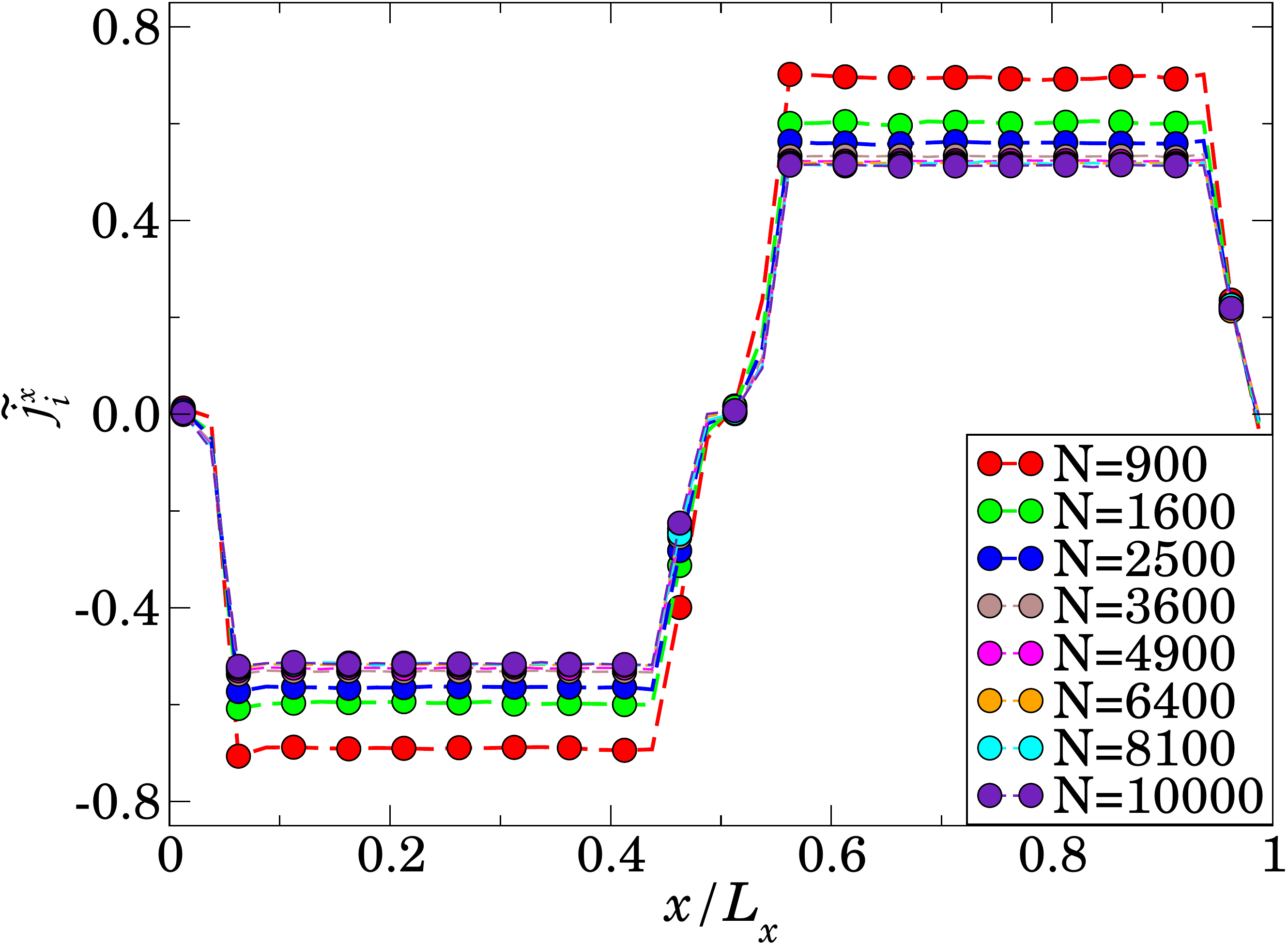}
\includegraphics[scale = 0.30]{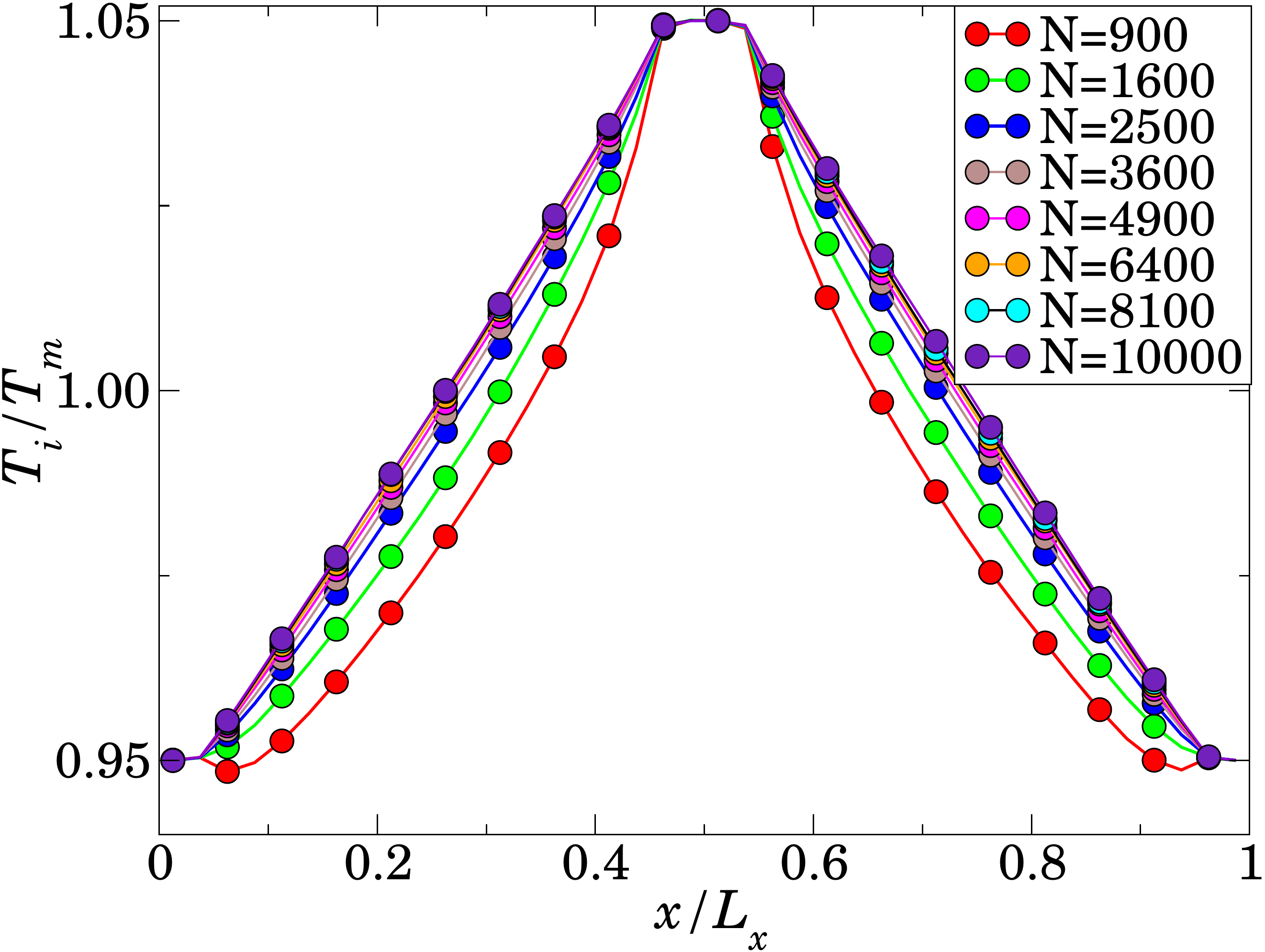}
\caption{(color online). (Top) $\tilde{\mathpzc{j}}^x_i$ and (bottom) reduced temperature($T_i/T_m$) profile in the $x$ direction for different system sizes at $T_m=0.5$. (Here, $T_m$ is the mean temperature of the sample in non-equilibrium simulation). The different systems have an equal length ($L_x=100a=94.1$) along the direction of temperature gradient, and varying cross sectional lengths $L_y(=L_z)=$: $ 3a(N=900),~4a(N=1600),~5a(N=2500),~6a(N=3600),~7a(N=4900),~8a(N=6400),~9a(N=8100)$ and $10a(N=10000)$.}
\label{figure_S1}
\end{figure}

{\em (a) Nonequilibrium simulation} -- In the nonequilibrium simulation we consider a rectangular box. The cross sectional lengths of the box are $L_y$ and $L_z$, with $L_y=L_z$.  Along the x-direction of the box, a temperature gradient is applied as described in the main paper. A heat source at temperature $1.05T_m$ is kept in a region of width $L_x/10$ in the middle of box while two heat sinks at temperature $0.95T_m$ each of width $L_x/20$ are kept at both ends of the box. The box is divided into $40$ segments, along $L_x$, labelled $i=1,\ldots40$. After an initial transient time, when the system reaches a non-equilibrium steady state, we calculate the local temperature $T_i$ and local heat current density $\langle \widetilde{\mathcal{J}}^x_i \rangle$ in each of the segments. While calculating these local quantities for a given segment we consider only the particles which are inside it, but the interaction of them will all other particles are taken into account. 

The bulk thermal conductivity($\kappa$) of the system is calculated from Eq.~(2) in the main paper, using the average heat current density $\langle \mathcal{J}_x \rangle$ measured over the volume ($v^\prime=L_x^\prime L_y L_z$) of the region of the sample across which the temperature gradient is imposed. We define the local quantity $\tilde{\mathpzc{j}}^x_i$ in terms of the  local heat current density $\langle \widetilde{\mathcal{J}}^x_i \rangle$ as,
\begin{equation}
\tilde{\mathpzc{j}}^x_i=\langle \widetilde{\mathcal{J}}^x_i \rangle \times L_x^\prime~,
\label{Eq:Lockappa}
\end{equation}
where, $L_x^\prime=0.4\times L_x$ is the length across which the thermal gradient is applied.
 
We have  performed a set of simulations to look into the effects of finite size of the system on $\kappa$ in non-equilibrium simulations at a fixed number density $\rho=1.2$. In the following we use $a=0.941\times\sigma_{AA}=0.941$ and express the size of the simulation boxes in units of $a$.
\begin{figure}
\includegraphics[scale = 0.3]{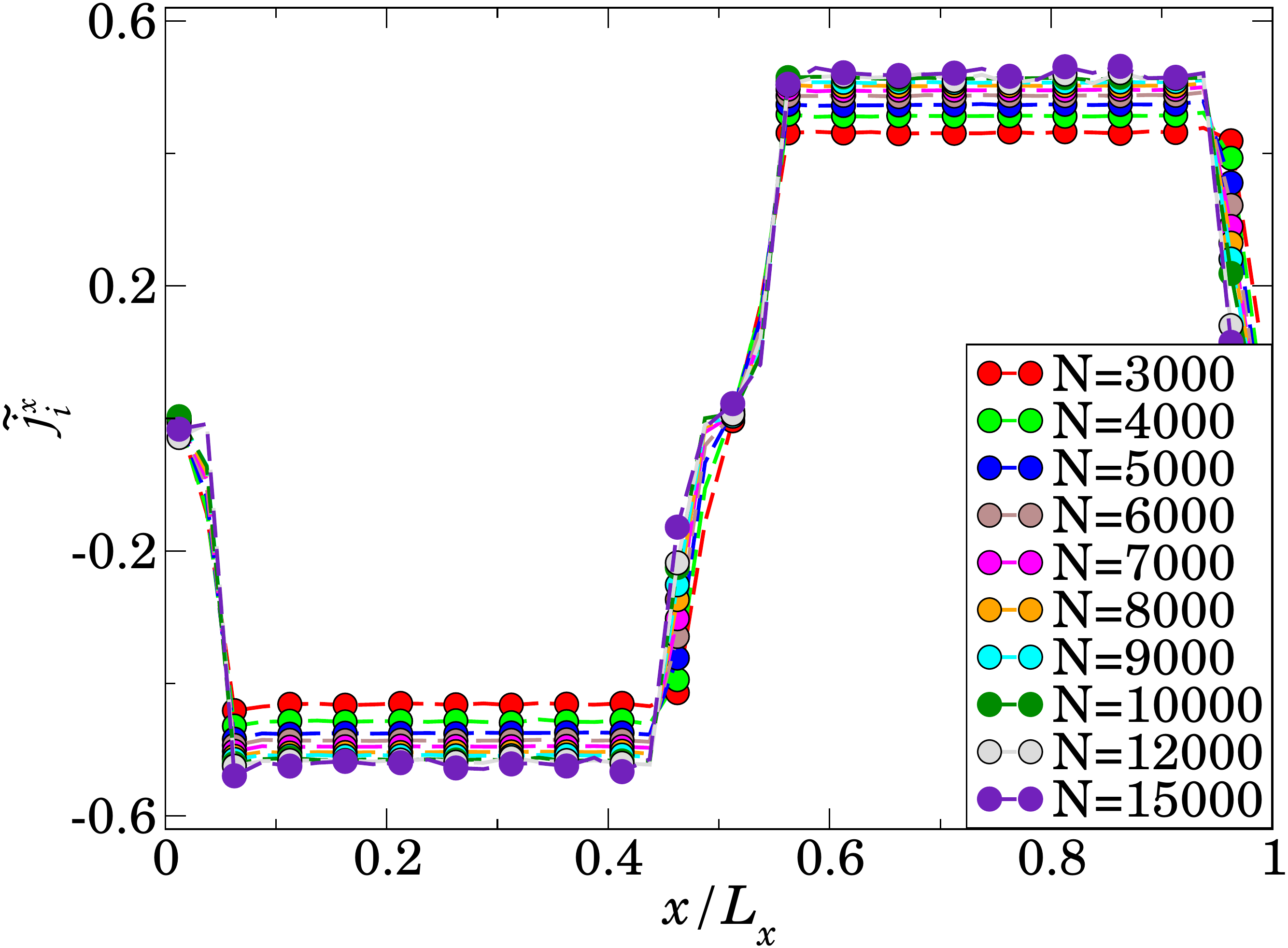}
\includegraphics[scale = 0.3]{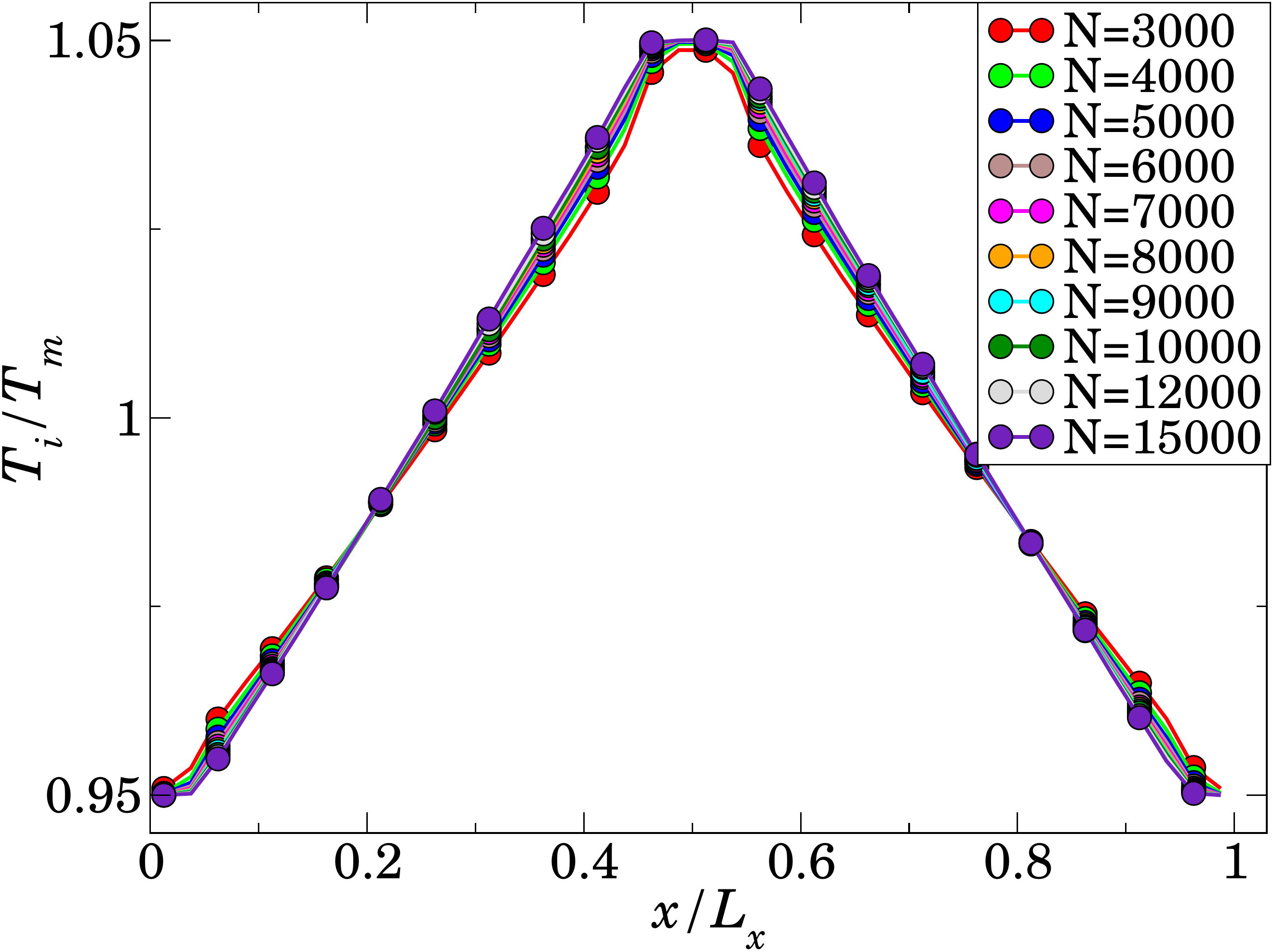}
\caption{(color online). (Top) $\tilde{\mathpzc{j}}^x_i$ and (bottom) reduced temperature ($T_i/T_m$) profile in the $x$-direction for different system sizes at $T_m=0.5$. (Here, $T_m$ is the mean temperature of the sample in non-equilibrium simulation). The different systems have equal cross-sectional lengths ($L_y=L_z=10a=9.41$), and varying lengths ($L_x$) along the direction of temperature gradient. $L_x=$: $30a(N=3000),~40a(N=4000),~50a(N=5000),~60a(N=6000),~70a(N=7000),~80a(N=8000),~90a(N=9000),~100a(N=10000),~120a(N=12000)$ and $150a(N=15000)$ .} 
\label{figure_S2}
\end{figure}

{\em Varying Cross sections} -- We use simulation boxes with a fixed length ($L_x=100a$) along the direction of temperature gradient and different cross-sections with $L_y=L_z$. In the top panel of Fig.~\ref{figure_S1},~ $\tilde{\mathpzc{j}}^x_i$ and in the bottom panel the temperature profile ($T_i/T_m$) along $L_x$ is shown.  We see that, for systems with $L_y(=L_z) \sim6a~(N=3600)$ finite size effects become negligible.

{\em  Varying Lengths} -- We fix the cross sectional lengths of the simulation box at $L_y=L_z=10a$ and use systems of different length $L_x$ along the directions of temperature gradient. The top panel of Fig.~\ref{figure_S2} shows $\tilde{\mathpzc{j}}^x_i$ and the bottom panel shows the temperature profile in the system. For systems with $L_x\sim80a~(N=8000)$ the system size effects become negligible.

\begin{figure} [ht]
\includegraphics[scale = 0.3]{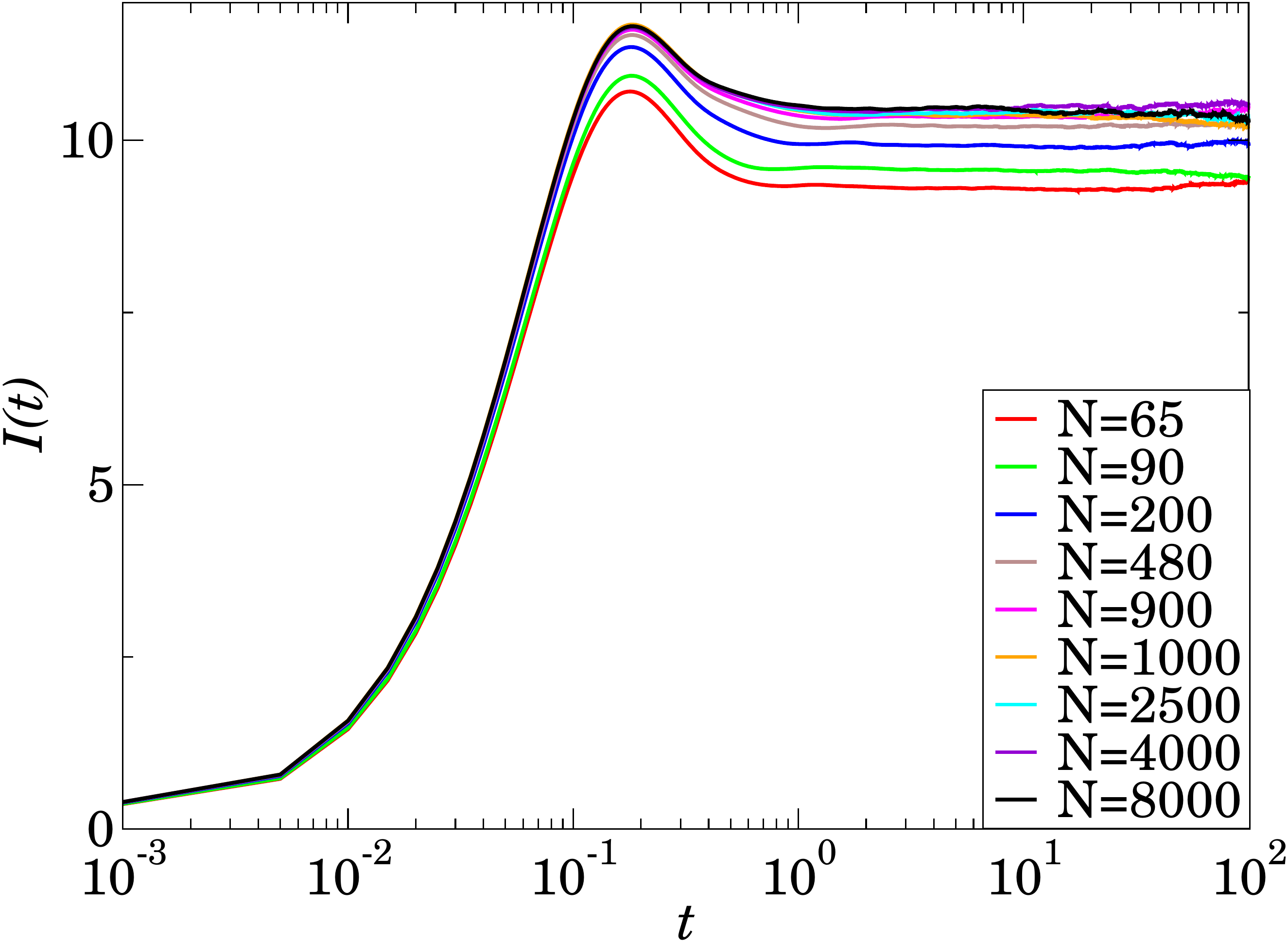}
\caption{(color online). $I(t)$ for different system sizes at $T=0.5$. The long time limit of $I(t)$ becomes independent of system sizes for $N\sim 900$.}
\label{figure_S3}
\end{figure}

{\em (b) Green-Kubo calculation} -- In the Green-Kubo calculation we use a cubic simulation box (i.e. $L_x=L_y=L_z$) and perform simulations for a number of system sizes. The conductivity $\kappa$ is calculated using Eq.~(3) as explained in the main paper. 

Following Eq.~(3) of the main paper we define:

\begin{equation}
I(t) = \frac{1}{3k_BT^2}  \frac{\rho}{N} \int_{0}^{t} dt^\prime \mspace{1mu} \langle \vec{\mathcal{J}}_{tot}(0) \cdotp \vec{\mathcal{J}}_{tot}(t^\prime)\rangle 
\label{Eq.integral}
\end{equation}

Fig.~\ref{figure_S3} shows the plots of $I(t)$ for different system sizes. The values of conductivity $\kappa$ are calculated by taking averages of the long time limit of $I(t)$ over a number of points in an interval between two time points where the it has a convergent value. For a simulation box with $N\sim 900$ finite size effects are found to be negligible.

\section{III. The harmonic approximation}

{\em (a) Heat diffusivity} -- The heat diffusivity of a normal mode with frequency $\omega_m$ and corresponding eigenfunction $\vec{e_i}(m)$ can be calculated as \cite{Allen:93,Nagel:09,Nagel:10}:
\begin{align}
d(\omega_m,\eta,N)\equiv & \frac{\pi}{12\omega_m^2}\sum_{n\neq m} \frac{(\omega_m+\omega_n)^2}{4\omega_m\omega_n}\times \nonumber \\
            & \lvert \vec{\Sigma}_{mn} \rvert^2 f(\omega_n-\omega_m,\eta)~,
\label{eqn.domega}
\end{align}
where $\vec{\Sigma}_{mn}$ is the heat flux matrix 
\begin{align}
\vec{\Sigma}_{mn} = \sum_{i,j,\alpha,\beta} (\vec{r}_i-\vec{r}_j) e_i^\alpha(m) H^{ij}_{\alpha \beta} e_j^\beta(n)
\label{eqn.sigma}
\end{align}
with $H^{ij}_{\alpha \beta}$ is the Hessian matrix elements and $\vec{r}_i$ is the position of the $i$-th particle; $\alpha, \beta=\left\{x,y,z\right\}$ are the components. The function $f(\omega_m-\omega_n;\eta)=
\eta/\{\pi[(\omega_m-\omega_n)^2+\eta^2]\}$ is a finite-width representation of the  Dirac-$\delta$-function that is necessary when dealing with finite systems. The width of the Lorentzian is chosen to be $\eta=\gamma \delta \omega$ with $\gamma > 1$, where $\delta \omega$ is the average spacing between successive modes.

\begin{figure} 
\includegraphics[scale = 0.3]{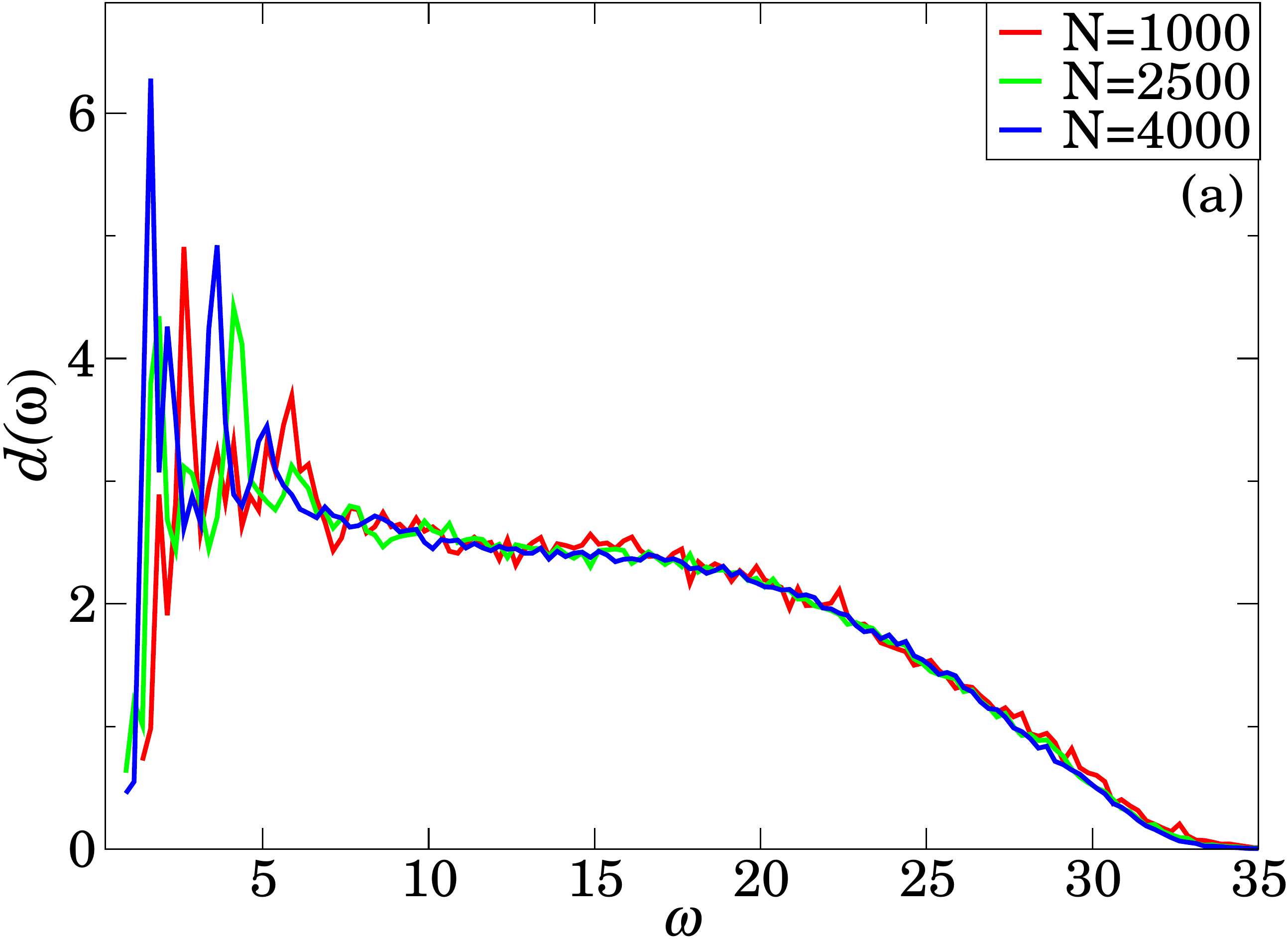}
\includegraphics[scale = 0.3]{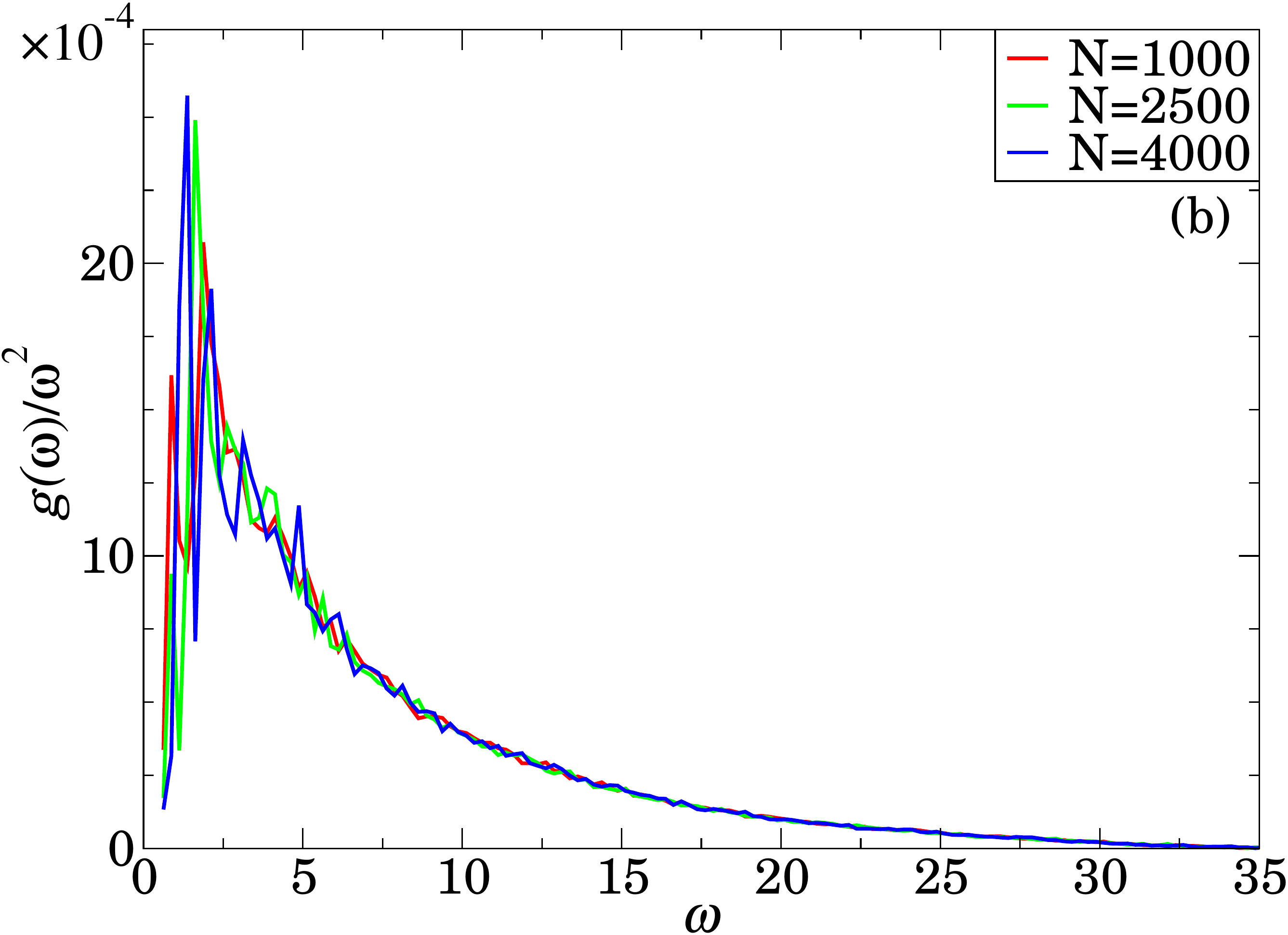}
\includegraphics[scale = 0.3]{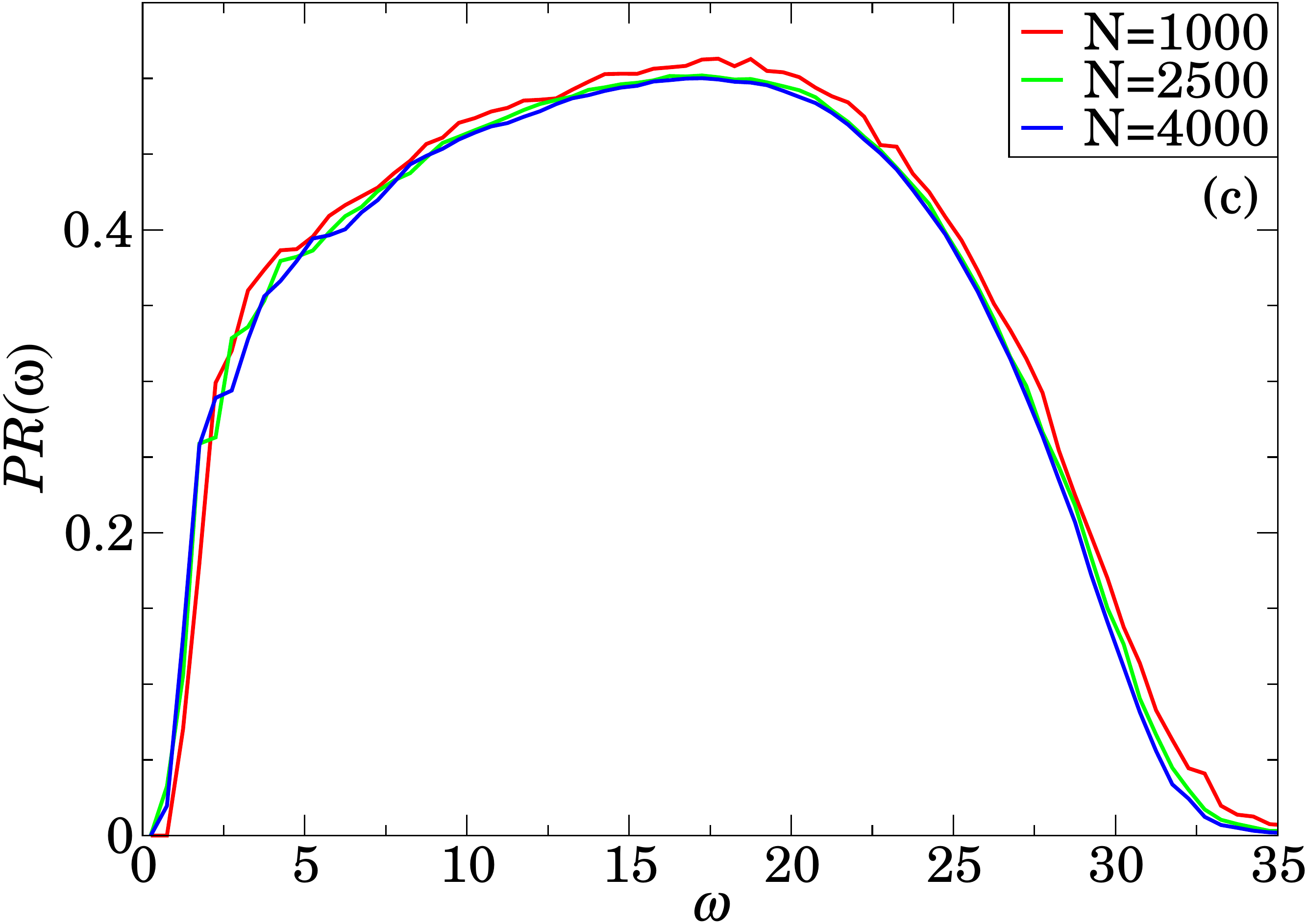}
\caption{(color online). (a) $d(\omega)$ for different system sizes (b) reduced density of states $g(\omega)/\omega^2$ and (c) participation ratio ($PR(\omega$)) for the same systems.}
\label{figure_S4}
\end{figure}

\vspace{1em}

{\em (b) Thermal Conductivity} -- The thermal conductivity $\kappa$ (from the Green-Kubo formula) for a harmonic system at temperature $T$ can be expressed in terms of the \textit{heat diffusivity d($\omega$)}~\cite{Nagel:09} as
\begin{align}
 \kappa=\frac{1}{V}\sum_m C(\omega_m) d(\omega_m)  
=\frac{1}{V}\int_{0}^{\infty} d\omega g(\omega)  C(\omega) d(\omega) ,
\label{eqn.kappa}
\end{align}
where $V$ is the volume of the system, $g(\omega)= \sum_n \delta(\omega -\omega_n)$ is the density of states, $C(\omega)= k_B {(\beta \hbar\omega)}^2 e^{\beta\hbar\omega}/{(e^{\beta\hbar\omega}-1)^2}$
is the phonon heat capacity, with $\beta \equiv 1/k_B T$ and $k_B$ is the Boltzmann constant. In our numerical calculations, we take $C(\omega)=k_B$, since we restrict our study to the classical domain.

\vspace{1em}

{\em (c) Participation Ratio} -- The participation ratio $PR(\omega)$ quantifies the localization properties of a normal mode. It is defined as ~\cite{laird91},

\begin{equation}
PR(\omega_n) =  \left[ N \sum\limits^N_{i=1} ( \vec{e}_i(n).\vec{e}_i(n) )^2 \right]^{-1} 
\label{eqn.PR}
\end{equation}

{\em (d) Finite size effects in the harmonic approximation} -- 
We take a few inherent structures of average energy $\overline{E}_{IS}$(=-7.0008) for a number of different system sizes and calculate thermal conductivity within the harmonic approximation using Eq.~\ref{eqn.kappa}. For all the calculations we use $\gamma=2.0$ for approximating the density of vibrational states in Eq.~\ref{eqn.domega}. Finite size effect on $\kappa$ is found to be insignificant for systems with $N\sim 1000$. 

In Fig.~\ref{figure_S4} we show the heat diffusivity (a) , boson peak (b) and the participation ratio (c) for these three different system sizes.

\vspace{1em}

{\em (e) Contribution of different parts of the eigenfrequency spectrum to $\kappa$ -}

\begin{figure}[h]
\includegraphics[scale = 0.30]{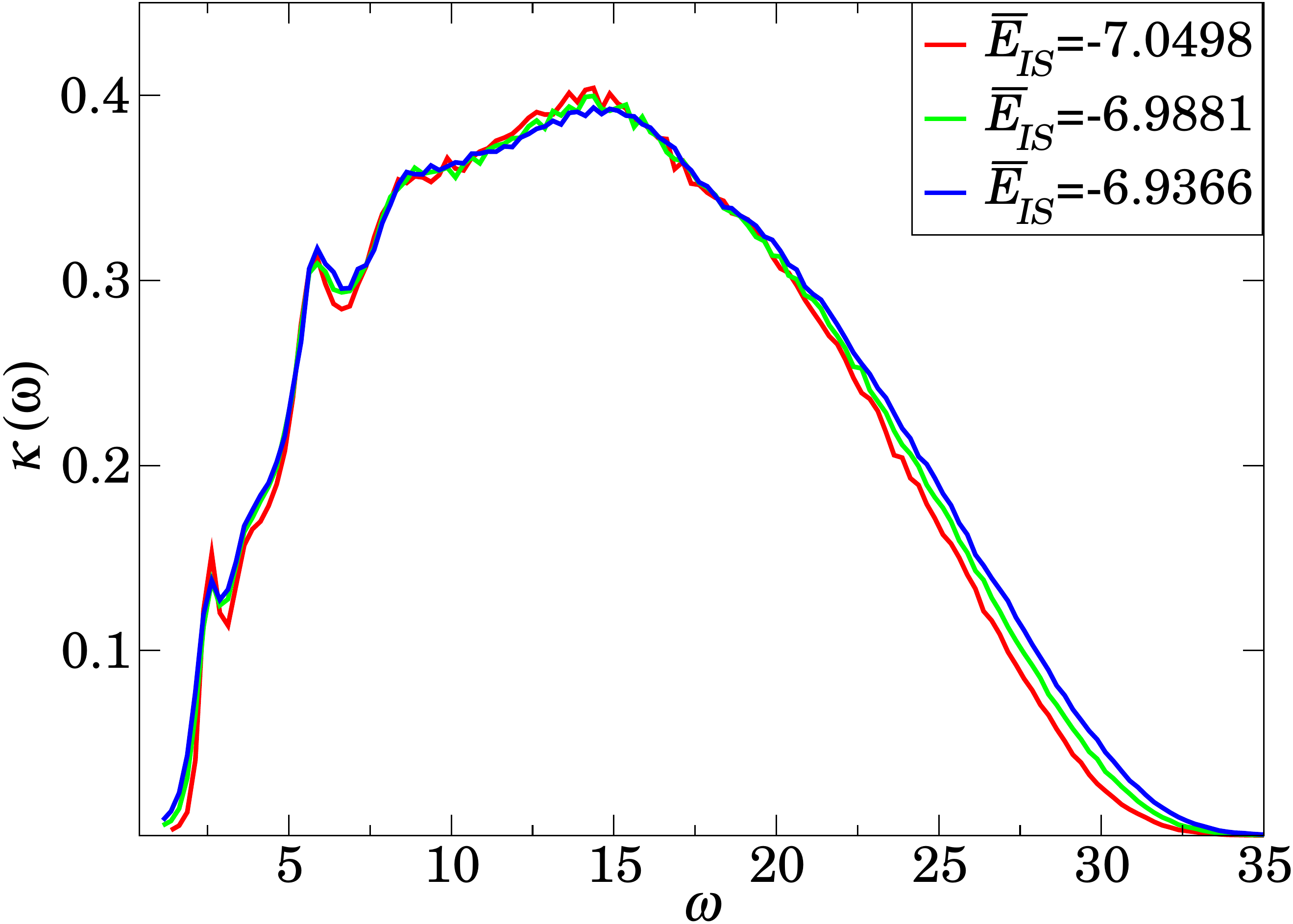}
\caption{(color online). $\kappa(\omega)$ for three different inherent structures.}
\label{figure_S6}
\end{figure}
Following Eq.~\ref{eqn.kappa} we define the thermal conductivity $\kappa$ for different frequencies as follows:
\begin{align}
 \kappa(\omega)=\frac{1}{V}~g(\omega)  C(\omega) d(\omega) 
\label{eqn.kappa.om}
\end{align}
with, $C(\omega)=k_B$ and $V$ is the volume of the system. We have calculated $\kappa(\omega)$ for the three groups of inherent structures with mean energy $\overline{E}_{IS}$. Fig.~\ref{figure_S6} shows the contribution of the different parts of the vibrational spectrum to the total $\kappa$. By performing the sum of $\kappa(\omega)$ over different frequency ranges we find that the largest contribution to $\kappa$ comes from the vibrational modes in the middle of the spectrum, where the modes are the most delocalized. However, the maximum contribution to the difference in the value of $\kappa$ between inherent structures of different energy are from the modes in the higher end of the vibrational spectrum.

{\em (g) Dependence of heat diffusivity $d(\omega)$ on the value $\gamma$} --

\begin{figure}[h]
\includegraphics[scale = 0.30]{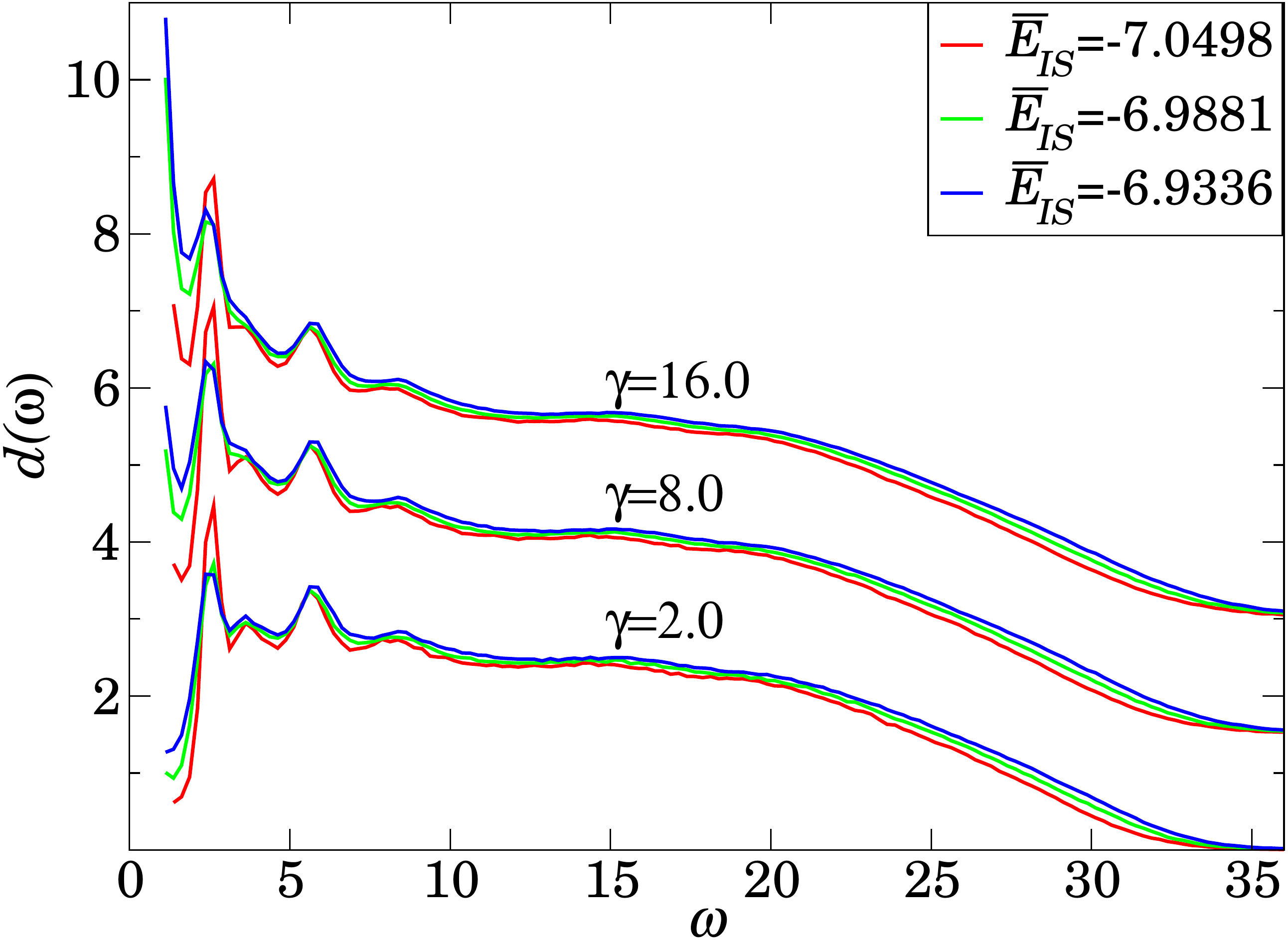}
\caption{(color online). Heat diffusivity $d(\omega)$ for three different groups of ISs with different $\overline{E}_{IS}$ for three different values of $\gamma$. For convenience of comparsion the values of $d(\omega)$ for $\gamma=8.0$ and $16$ has been shifted by 1.5 and 3.0 respectively.}
\label{figure_S7}
\end{figure}

 The values of heat diffusivity, $d(\omega)$, calculated from the harmonic calculation is sensitive to the choice of $\gamma$ (the width of the broadening function). In Fig.~\ref{figure_S7} we show the dependence of $d(\omega)$ on $\gamma$ for the three different groups of IS shown in Fig.6 of the main paper. The nature of the dependence of $d(\omega)$ on the IS energy does not change with changing $\gamma$ for most values of $\omega$, except at the low frequency end where numerical artifact starts appearing because of the overestimation in the desnity of states for higher values of $\gamma$.

{\em (h) Comparison of $\kappa$ values from Non-equilibrium simulation and harmonic calculation} --

\begin{figure}[ht]
\includegraphics[scale = 0.30]{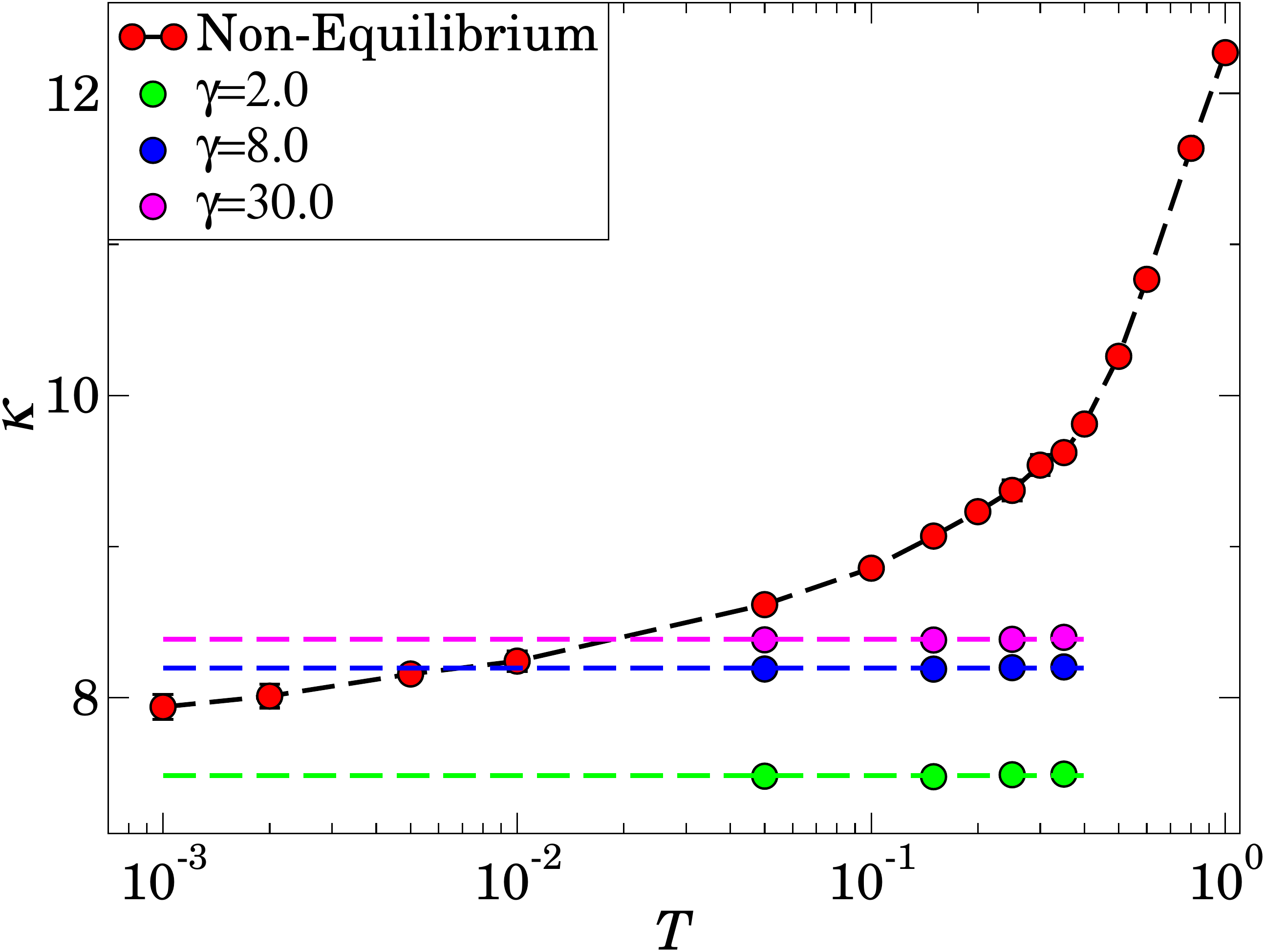}
\caption{(color online). Thermal conductivity ($\kappa$) for different temperatures from non-equilibrium simulation at cooling rate $3.33\times10^{-6}$. The horizontal dashed lines along with the points show the values of $\kappa$ calculated in harmonic approximation, sampling ISs at few low temperatures, for three different values of $\gamma$.}
\label{figure_S8}
\end{figure}

As illustrated in section (g) above the values of $d(\omega)$ and consequently the thermal conductivity, $\kappa$, calculated from the harmonic calculation is sensitive to the choice of $\gamma$. To obtain the true $\kappa$, one needs to take the $N \to \infty$, $\eta \to 0$ limit. For finite systems, the choice of $\gamma$ leads to some uncertainty in the calculated $\kappa$. In Fig.~\ref{figure_S8} we compare the values of $\kappa$ at low temperatures obtained from non-equilibrium simulation to those obtained from harmonic calculation performed with three different choices of $\gamma$ from ISs derived at few low temperatures. 


\section{IV. Direct numerical verification of the harmonic approximation at low temperatures} 

\begin{figure} 
\vspace{2em}
\includegraphics[scale = 0.30]{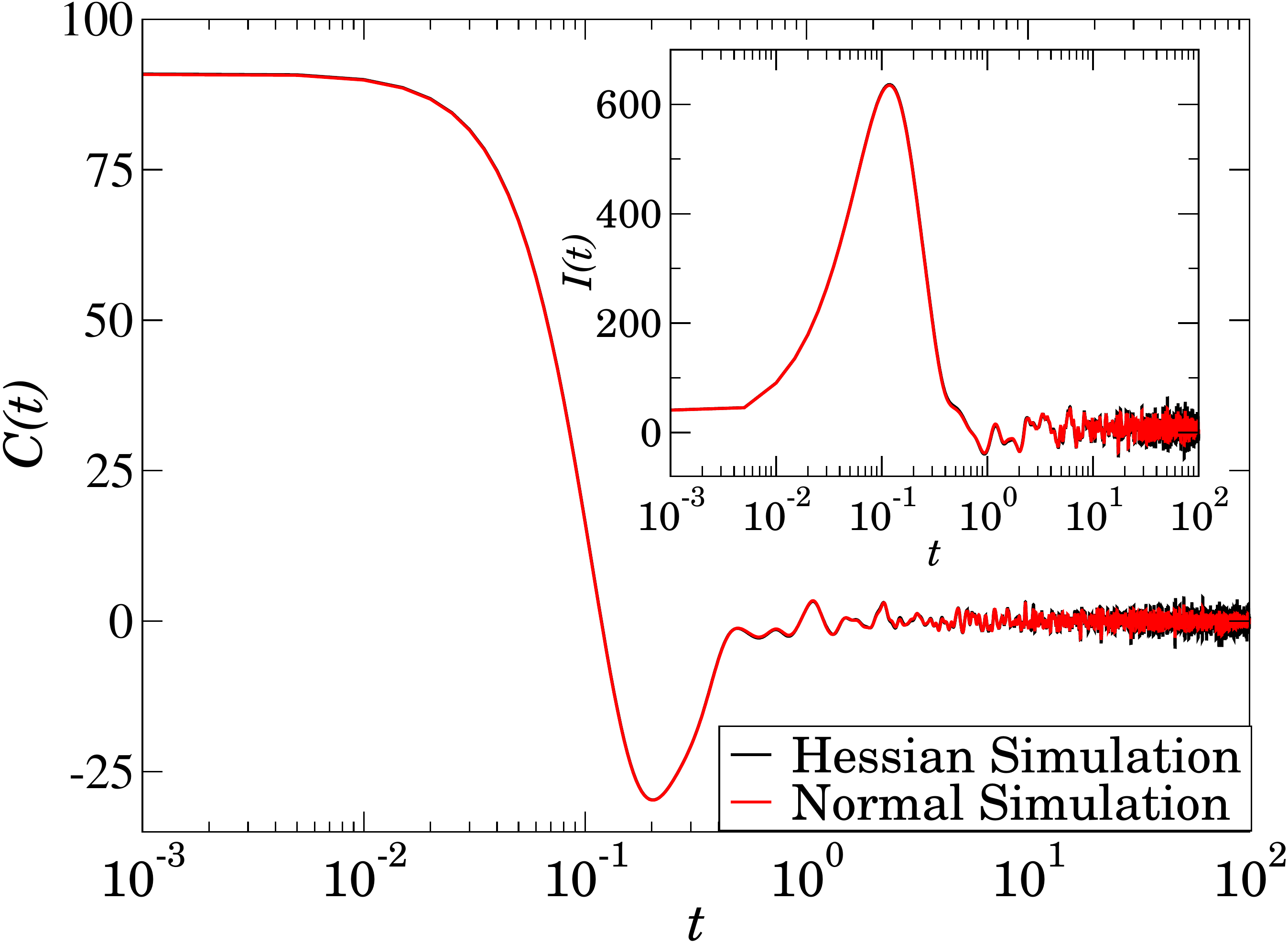}
\caption{(color online). Current autocorrelation function C(t) and its integral I(t), shown in the inset, calculated from normal and Hessian simulations.}
\label{figure_S5}
\end{figure}

One can ask whether at low temperatures the system actually stays within the basin of a given inherent structure (a local minimum of the potential) and also if nonlinear effects are important for dynamics within the basin of an inherent structure. 
Here we examine this question of the validity of the harmonic approximation at low temperatures.
We computed and compared the energy current auto-correlation function  $C(t)= \langle \vec{\mathcal{J}}(0) \cdotp \vec{\mathcal{J}}(t)\rangle$, at $T=0.002$, using molecular dynamics with the actual interactions and within the harmonic approximation for an inherent structure.  

We prepare an initial state by choosing positions corresponding to an IS and giving each particle a random velocity taken from a Maxwell distribution at temperature $2T$. Within a short span of time the energy is equipartioned. With these initial conditions we first perform MD simulations with the full inter-particle potential (Normal Simulation) and secondly, with the inter-particle forces replaced by the forces from the Hessian elements (Hessian Simulation). We consider only those low energy ISs for which 
the system stays inside the initial basin during the course of the simulation. The results are shown in Fig.~(\ref{figure_S5}). We see excellent agreement between the two which implies that the harmonic approximation is indeed quite accurate at this temperature.
